# PRODUCTIVITY OF PRE-MODERN AGRICULTURE IN THE CUCUTENI–TRYPILLIA AREA

A. Shukurov[a], M. Y. Videiko[b], G. R. Sarson[a], K. Davison[a], R. Shiel[c], P. M. Dolukhanov[d,e] and G. A. Pashkevich[b]

[a] *School of Mathematics and Statistics, Newcastle University, Newcastle upon Tyne, NE1 7RU, UK*
[b] *Institute of Archaeology, National Academy of Sciences of Ukraine, Kyiv, 04210, Ukraine*
[c] *School of Agriculture, Food and Rural Development, Newcastle University, Newcastle upon Tyne, NE1 7RU, UK*
[d] *School of History, Classics and Archaeology, Newcastle University, Newcastle upon Tyne, NE1 7RU, UK*
[e] *Deseased*

## Abstract

We present palaeoeconomy reconstructions for pre-modern agriculture; we select, wherever required, features and parameter values specific for the Cucuteni–Trypillia Cultural unity (CTU: 5,400–2700 BC, mostly the territory of modern Ukraine, Moldova and Romania). We verify the self-consistency and viability of the archaeological evidence related to all major elements of the agricultural production cycle within the constraints provided by environmental and technological considerations. The starting point of our analysis is the palaeodiet structure suggested by archaeological data, stable isotope analyses of human remains, and palynology studies in the CTU area. We allow for the archeologically attested contributions of domesticated and wild animal products to the diet, develop plausible estimates of the yield of ancient cereal varieties cultivated with ancient techniques, and quantify the yield dependence on the time after initial planting and on rainfall (as a climate proxy). Our conclusions involve analysis of the labour costs of various seasonal parts of the agricultural cycle of both an individual and a family with a majority of members that do not engage in productive activities. Finally, we put our results into the context of the exploitation territory and catchment analysis, to project various subsistence strategies into the exploitation territory of a farming settlement.

The simplest economic complex based on cereals, domestic and wild animal products, with fallow cropping, appears to be capable of supporting an isolated, relatively small farming community of 50–300 people (2–10 ha in area) even without recourse to technological improvements such as the use of manure fertiliser. Our results strongly suggest that dairy products played a significant role in the dietary and labour balance. The smaller settlements are typical of the earliest Trypillia A but remain predominant at the later stages. A larger settlement of several hundred people could function in isolation, perhaps with a larger fraction of cereals in the diet, only with technological innovations, such as manure fertiliser and, most importantly, ard tillage. The ard relieves radically the extreme time pressure associated with soil preparation for sowing. It appears that very large settlements of a few hundred hectares in area could function only if supported by satellite farming villages and stable exchange networks. In turn, this implies social division of labour and occupation, sufficiently complex social relations, stable exchange channels, etc.: altogether, a proto-urban character of such settlements. We also discuss, quantify and assess some strategies to mitigate the risks of arable agriculture associated with strong temporal fluctuations in the cereal yield, such as manure fertilisation, increased fraction of cereals in the diet combined with producing grain surplus for emergency storage.

# 1. Introduction

The economy and demography of the spread and subsequent development of early agriculture, and their mathematical modelling, remain one of the predominant themes in the studies of prehistory. Most previous work on the mathematical modelling has focused on the 'first arrival' of the Neolithic (see Steele 2009 and Fort 2009 for a review). Here we attempt to provide a quantitative basis for the essentially nonlinear modelling of the subsequent evolution of the farming population in a newly colonized area. Among the relevant processes, some of them identifiable from archaeological and radiometric evidence, are the evolution of the population density after the initial settlement stage, the spatial clustering of the population, as well as the development of (hierarchical) settlement patterns and exchange and communication networks.

Models of the initial spread of the Neolithic involve a number of parameters mainly estimated from ethnographic and archaeological evidence. These include the intrinsic growth rate of the population, its mobility (or diffusivity), and the carrying capacity of the landscape. An important aspect of carrying capacity estimations is the productivity of early farming, including its dependence on major environmental parameters. Our subject here is palaeoeconomy reconstructions that underpin carrying capacity estimates. We verify our results by comparing the resulting maximum size and lifetime of a farming settlement with the archaeological data for the Cucuteni–Trypillia culture (ca 5,400–2,700 BC, the territory of modern Ukraine, Moldova and Romania).

Apart from generic data on agricultural productivity, our estimates of the cereal yield, and its dependence on climate and soil depletion, are derived using data from an experimental agricultural farm in the US Midwest, where the climate and soil type are broadly similar to those in the Cucuteni–Trypillia area. The data suggest that, for a given soil type and crop variety, the January–May rainfall, the use of natural fertilisers, and the cultivation time are the main variables that control the yield produced. In what follows, we quantify the dependence of the wheat crop yield on these variables, and proceed to including animal husbandry and diet variations into our model of the productivity of pre-modern agriculture.

Palaeoeconomy reconstructions for early agricultural communities are numerous and diverse, at both global and regional levels (Higgs and Vita-Finzi 1972; Jarman et al. 1982; Ellen 1982; Gregg 1988; Ebersbach and Schade 2004; Tipping et al. 2009). There is a number of such studies for the CTU agriculture in particular (Bibikov 1964; Krutz 1989; Zbenovich 1996; Nikolova and Pashkevich 2003; Videiko et al. 2004; Pashkevich and Videiko 2006). Many such studies aim, explicitly or implicitly, to estimate the carrying capacity of the landscape. However, it is impossible to disagree with the opinion expressed in Jarman et al. (1982, p. 24), that "the production of precise numerical population estimates" is "a most hazardous undertaking given the uncertainty surrounding resource levels… One tends thus to be faced with a figure so hedged about with qualifications, or so slenderly justified, as to command little confidence". Indeed, the usefulness of such calculations is not in the resulting figures, even though they must be of a reasonable magnitude and consistent with other relevant knowledge to be acceptable. Any estimates of this kind cannot be used to assess, even in rough terms, the population of any region. Their significance is rather in (a) an opportunity of quantitative hypothesis testing; (b) confirmation (or otherwise) of the mutual consistency of various elements of the overall palaeoeconomy and subsistence picture and, most importantly, (c) assessment of the effect of the input parameters and identification of the most important of them, that is those to which the results are most sensitive. Such parameters should be the first to attract further attention as to obtain their reliable values. Furthermore, results based on the same principles but applied to different regions or even epochs can help to assess their relative similarities and dissimilarities.

Quoting Jarman et al. (1982, p. 14) again, "Man (along with pigs and rats), however, is dietarily an omnivore… Thus the computation of human nutritional requirements is immensely complicated". Indeed, calculations presented here involve a large number of parameters. The values of many of them in the context of early agriculture are known poorly or unknown. There-fore, a large part of our effort was to collect and summarize relevant data, translate them to the prehistoric context if required, and then to isolate results that are less dependent on hypothetical constructs. We focus on the sole characteristic of the food production system, the calorific value of cereals and meat products leaving aside numerous other components of the economic and social system. We are far, however, from suggesting that the simplest constraints that can be identified with this approach are predictive and deterministic. But as Ellen (1982, p. 123) notes, "Much of what we say about the operation of specific social systems must hinge on an accurate appreciation of how social relations articulate with pattern and techniques of subsistence". Our aim here is to contribute to a quantitative understanding of the "pattern and techniques of subsistence" of the Neolithic and Bronze Age farmers, those in the CTU area in particular.

Our attempt at the palaeoeconomy reconstruction is somewhat different from the earlier approaches. We first establish a set of plausible estimates of the numerous important parameters that characterize early farming (both plant and animal husbandry), verify that they are not self-contradictory by assessing the land use and labour costs of the agricultural production consistent with them, and then discuss the dependence of the economic behaviour of the population on various input parameters and their combinations. The last step allows us to isolate robust results and separate those factors that affect the farming economy most profoundly and thus warrant further archaeological investigation. We deliberately neglect a large number of details in our models and calculations (such as the difference between calorific values of various cereal varieties grown by the CTU farmers, the difference in the calorific content of hay and leafy fodder, etc.) retaining only those parameters that can affect the results rather dramatically. Firstly, many of such details are subsumed into gross features that, unlike the details, can be quantified using archaeological, environmental and ethnographic evidence. Secondly, excessive details (which are not, in fact, difficult to include) can lead to an illusion of a high precision, accuracy and predictive power of the results, which are unavoidably very limited in such calculations.

## 2. The Cucuteni–Trypillia cultural complex

One of the most important and best-explored early farming communities in Eastern Europe is the Late Neolithic–Chalcolithic Cucuteni–Trypillia cultural unity (CTU). Discovered independently in eastern Romania (Cucuteni) and in the central-western Ukraine (Trypillia) in the late 19$^{th}$ century, the CTU underwent several stages in its evolution specified in Table 1. Extensive reviews of the nature and development of the CTU, briefly summarised here, can be found in Zbenovich (1996) and Videiko (2000). A comprehensive review of the CTU archaeology is presented in Videiko et al. (2004).

Table 1. Chronology of the Cucuteni–Trypillia Unity (Videiko 2003; Klochko and Krutz 1999; Kovalyukh et al. 1996)

| Cultural Stage | | Time Span, years BC |
|---|---|---|
| In Ukraine | In Romania | |
| Trypillia A | Precucuteni I, II, III | 5400/5300–4800/4700 |
| Trypillia BI | Cucuteni A (1–4) | 4800/4700–4500/4400 |
| Trypillia BI/II | Cucuteni A-B (1–2) | 4500/4400–4100/4000 |
| Trypillia BII + CI | Cucuteni B (1–3) | 4100/4000–3400/3300 |
| Trypillia CII-γII | Gorodiştea–Folteşti–Erbiceni | 3400/3300–2800/2700 |

CTU sites are located either in close proximity to, or within, river valleys, in most cases on natural elevations. The number of CTU sites found in the territory of Ukraine alone is about 2,100; most of them are permanent settlements. Table 2 presents the areas of the sites. The typical (median) area of Trypilla settlements is significantly smaller than their mean area at each stage because there is a relatively small number of exceptionally large settlements that affect the average but not the median area. The difference between the mean and the median areas is not very strong at the earlier stages A–BI but becomes extreme at the later stages. In such cases, the median area best represents a typical site. There is a systematic increase in the size of the settlements, with a maximum during the middle stages.

Table 2. The mean, median and maximum areas of CTU sites in the Ukraine per stage, and the number of sites with known area

| Stage           | A   | BI  | BI–BII | BII  | BII–CI | CI   | CII |
|-----------------|-----|-----|--------|------|--------|------|-----|
| Mean area, ha   | 3.0 | 9.9 | 28.8   | 14.6 | 12.1   | 20.2 | 9.1 |
| Median area, ha | 2.0 | 5.0 | 6.4    | 2.0  | 8.4    | 6.0  | 1.8 |
| Maximum area, ha| 14  | 60  | 150    | 261  | 150    | 341  | 160 |
| Number of sites | 20  | 15  | 28     | 119  | 53     | 151  | 56  |

Plant remains identified at the CTU sites in the Ukraine and Moldova show that agriculture was already substantially advanced, even at early CTU stages. The dominant species of cereals were hulled wheats (*Triticum dicoccum* Schrank*, T. monococcum* L. and *T. spelta* L.*)*, supplemented by naked six-row barley (*Hordeum vulgare* var. *nudum* Hook f.*coeleste* L.) and hulled barley (*Hordeum vulgare*). Broomcorn millet (*Panicum miliaceum* L.) was less common. During later periods, changes are observable only in the dominant varieties of barley: large amounts of naked barley were particularly typical of Trypillia A/Precucuteni sites, but were increasingly replaced by hulled varieties. The list of Trypillia cultigens also included pea (*Pisum sativum* L.) and bitter vetch (*Vicia ervilia* L.); pulse seeds are also frequently recovered in excavations. The fields were cultivated with antler and stone hoes, which made the soil more friable and thus better prepared for sowing the spikelets of hulled wheats. The use of the ard is suggested both by a find of an antler ard at Grebenukiv Yar (Pashkevich and Videiko 2006) and by cattle bone structures that suggest their use for traction (Zhuravlev 2008). The harvesting technique was probably specially adapted for cutting ears. Low yields, long periods of natural soil regeneration, primitive tools for soil cultivation and harvesting, and the use of undemanding cultigens were the basic features of the Early- and Middle-Trypillia agriculture.

The animal remains identified at the Trypillia sites belong to both wild species (red deer, wild boar, roe deer, elk, etc.) and domesticated species (cattle, pig, sheep/goat and horse); the relative occurrence of species varies significantly from site to site, implying considerable variations in subsistence. Cattle (and possibly horses) were used for transportation and traction as evidenced by bone structures and pottery models of sledges with ox heads found at several sites.

From the early phases, CTU settlements consisted of several one- or two-storey houses, each supposedly inhabited by a single family (sometimes, several families). The population of a typical settlement (estimated to be 50 to 500 people) formed a basic community unit, apparently sharing the ownership of land and other resources. No communal cemeteries are known at the CTU sites from the early and middle periods. From the earliest periods onwards, female effigies were predominant among the portable figurines, possibly symbols of fecundity, as grains of wheat and barley were found included in the ceramic fabric of several figurines at the Luka-Vrublevetskaya site (Bibikov 1953).

There are at least two concepts concerning the origins and expansion of the CTU; in the main, it is viewed as a result of migration from west to east and south. A different viewpoint, particularly popular in the former Soviet Union, stressed the local origin of the CTU, pointing to the Bug–Dniesterian region as the most likely source. Based on the bulk of available evidence one may consider the initial emergence of CTU sites in the forest-steppe of Eastern Europe as an agricultural colonization, essentially similar to that of the LBK in central Europe, with a complete culture-economic package spreading into a poorly occupied niche at a rapid pace. Similarly to the LBK, a limited impact of indigenous (in the CTU case, the Bug–Dniester) groups is rec-ognizable in the location of the sites and in the material culture. More recently, the possible influence on the CTU of agricultural innovations originating further east (e.g., proso millet, hemp), and migrating west via the 'Caucasus corridor', has received more serious attention (Motuzaite-Matuzeviciute et al., 2009).

CTU communities never existed in isolation; their extensive connections with neighbouring groups are recognizable in various aspects of their material culture. Contact with the East became particularly apparent during the middle phase, when the settlements expanded further eastward and grew in size. Several sites became particularly large: Vesely Kut reached 150 ha in size; Talyanky was still larger at 341 ha and had approximately 14,000 inhabitants (Pashkevich and Videiko 2006); the area of Maydanetske was 210 ha, with 2,900 houses identified by geophysical surveying. All these settlements were surrounded by fortifications consisting of palisades and houses built next to each other. At this stage, the Trypillia sites show signs of a growing social hierarchy, primarily evident in the occurrence of élite burials. The earliest recognizable kurgan-type barrow has been found in Moldova, at the site of Kainari. It contained a female skeleton with a rich collection of grave goods consisting of ceramic vessels (Trypillia BI) and copper adornments. Several Middle Trypillia sites included stone anthropomorphic sceptres and mace heads.

At that time, several distinct cultural entities arose in the steppe to the east of the Trypillia core area, including the Seredni Stig and Mykhailivka (Lower Level) cultures. The unfortified dwelling sites and cemeteries of Seredni Stig were located in forested river valleys, between the lower Dnieper and Don. The apparent distinctions in subsistence from the Trypillia area are primarily attributable to the ecology: an increasing aridity of the climate towards the east makes agriculture in the areas east of the Dnieper less sustainable and less predictable. One may reasonably suggest that, because of the increasing scarcity of water supply in the areas east of the Dnieper, the agricultural activities predominantly took the character of stockbreeding.

The area of Trypillia settlements lies in a temperately continental climatic zone influenced by moderately warm, humid air from the Atlantic Ocean (e.g., Pashkevich and Videiko 2006, p. 15). Winters in the west of the region are considerably milder than in the east, but the eastern part often experiences higher summer temperatures. Average annual temperatures range from 5.5–7° in the north to 11–13°C in the south. The average temperature in January, the coldest month, is −3°C in the southwest and −8°C in the northeast. The average temperature in July, the hottest month, is 23°C in the southeast and 18°C in the northwest.

Maximum precipitation generally occurs in June and July, while the minimum falls to February. The precipitation in the western part of the CTU area is 650 mm/year and decreases to 450–600 mm/year in the east. Western Ukraine, notably the Carpathian Mountains area, receives the highest annual precipitation of more than 1,200 mm/year. Snow falls mainly in late November and early December, varying in depth from 5–10 cm in the steppe region to several feet in the Carpathians. The snow cover in the Dniester–Prut interfluve is unstable but it can stay for up to 40 days in the eastern region.

Trypillia settlements are located in the area of fertile chernozem soils. The most fertile varieties, the so-called deep chernozems, lie in the north (about 1.5 m thick and rich in humus). Prairie, or ordinary, chernozems, equally rich in humus but only about 1 m thick, occur further south and east. The soil in the southernmost belt has an even thinner chernozem layer and has still less humus. Interspersed in the uplands and along the northern and western perimeter of the deep chernozems are mixed grey forest and podzolic black-earth soils, which together form the remaining soil cover. All these soils are very fertile when sufficient water is available. The smallest proportion of the soil cover consists of the chestnut soils of the southern and eastern regions, which become increasingly salinized to the south closer to the Black Sea.

# 3. Palaeodiet reconstructions

The relative importance of plant food versus domestic animal products and wild meat in the diet of early farming communities remains a subject of active discussion. Stable isotope analysis of human bones by Lösch et al. (2006) suggests that, in the early farming communities of Anatolia (Pre-Pottery Neolithic B, mid-ninth millennium BC), "the contribution of stock on the hoof in the human diet was modest". Low $^{15}$N values in their samples imply the increased consumption of protein-rich cereals and pulses. According to these authors, animal husbandry gained in importance at later Neolithic stages. Bogaard (2004a) concluded, from archaeobotanical evidence, that cereals and pulses provided the bulk of the diet in Neolithic Greece, while livestock provided a vital alternative in the case of crop failure.

In contrast, investigations of Copper Age (early- to mid-fifth millennium BC) cemeteries in Varna I and Durankulak, Bulgaria (Honch et al. 2006), using stable carbon ($^{13}$C/$^{12}$C) and nitrogen ($^{15}$N/$^{14}$N) isotope ratios, suggest a diet based on terrestrial resources, with a predominance of animal products (meat and/or milk, cheese and other secondary products from sheep/goat). These sites are roughly coeval with Trypillia A. However, the Bulgarian Copper Age sites are more advanced agriculturally. Hence, one might argue that the initial stage of farming at the early Trypillia sites may be structurally closer to the early Anatolian farming with the human diet being essentially based on cereals and pulses, with greater impact of animal husbandry at the later stages.

Ogrinc and Budja (2005) perform a similar stable isotope analysis of the animal (both wild and domestic) and human bone collagen as well as of floral remains (mostly wheat, barley and peas) from Ajdovska Jama cave in Slovenia, dated to 6400–5300 years cal BP, i.e., coeval with Trypillia B–C. These authors find convincing evidence for a stable palaeoeconomy during this whole period, based on terrestrial food resources. According to these results, the major diet components were domestic animal products (44%), cereals (39%) and terrestrial wild meat (17%). Bogaard et al. (2007) stress that field manuring can bias the results of such analyses, leading to an overestimation of the contribution of animal products to the diet. However, there is firm archaeological evidence in favour of the importance of animal husbandry in the CTU agriculture. Pashkevich (1989, p. 136) concludes, from palynology data, that land farming and animal husbandry were equally important at the Maydanetske settlement.

As a plausible estimate and the starting point of our discussion, we assume that domestic animal products and cereals provided each 40% of the food consumption of the CTU population, with the remaining 20% coming from hunting. The meat weight and its calorific value of the hunted animals (mostly red deer, roe deer and wild boar in Trypillia) can be found in Jarman et al. (1982, p. 83). We do not include vegetables and other plants in our calculations as they could only contribute little to the calorific content of the diet: although as much as 2–3 kg of leafy vegetables would supply as little as 1000 kcal of energy (Jarman et al. 1982, p. 16), this volume of food exceeds the natural biological constraints of the human body. Likewise, we do not include any wild plants even if their calorific value might be comparable to that of cereals (Stokes and Rowley-Conwy 2002).

Our calculations presented below refer to the energy content of the food alone, but not to any nutritional balance of its individual components such as proteins, vitamins, amino acids, etc. Moreover, we only consider cereals, meat and dairy products but neglect legumes. Jarman et al. (1982, p. 16) note that, "when adequate calories are available from a varied diet, then considerably more than minimal protein requirements are automatically provided". Given the unavoidably tentative and approximate character of palaeoeconomy calculations, we do not feel that introducing a more detailed nutritional classification of foods would be justifiable.

# 4. Cereal yield

In this section, we discuss methods of estimating the plausible wheat yield in the CTU region using the available data from agricultural experiments in other comparable areas. Apart from corrections for ancient wheat varieties, we present evidence for the variation of the yield with rainfall, duration of continuous cropping and the efficiency of manure fertilisation. Since no evidence of irrigation has been discovered in the CTU area, we focus on dry farming.

## 4.1 Agricultural experiments

Any attempt to estimate the productivity of prehistoric agriculture faces a number of problems. Specifically in the CTU area, the land in the Ukraine, Moldova and Romania today has been cultivated for 9000–8000 years and the soils are unlikely at all to have properties like those encountered by the first CTU farmers. The varieties of wheat grown today have been modified by plant breeders and the yields have increased greatly, even without fertilisers (Austin et al. 1993). Furthermore, agricultural tools have changed over time, undoubtedly affecting the agricultural productivity. Added to this is the problem that, in the modern agricultural practice and in most agricultural experiments, the soil is amended with nutrients and pests, often made heavier by prolonged use of heavy agricultural machinery, and with weeds and diseases controlled using synthetic chemicals.

One way to address a part of these problems is to use the results from long-term agricultural experiments in areas which had not previously been used for agriculture. This excludes virtually the whole of Europe, Africa and Asia. In the central United States, however, there are areas that are climatically similar to the Ukraine and where the prairies remained uncultivated until the late nineteenth century. In Australia, there are also similar areas that had not been exploited; however, in southern Australia, unlike the CTU area, the climate is Mediterranean with a severe summer drought. These experiments mostly involve modern wheat varieties rather than those used in the early agriculture. This remains a problem which is hard to resolve completely (see below).

Our main data come from the Sanborn Field of the Agricultural Experiment Station of the University of Missouri–Columbia, USA (N38°57′, W92°19′) which began in 1888 and still continues; this is one of the oldest continuous, long-term research plots in the world (Miller and Hudelson 1921). In this experiment, we are interested in wheat grown annually and in various biennial and rotational systems both with and without the use of manure fertiliser. The Sanborn field is divided into 39 experimental plots, each 30 m by 10 m in size, separated by 1.5 m wide grass hedges. Changes were made to the experiment over its lifetime. Commercial fertiliser was introduced in 1914, and the number of plots receiving manure was reduced, which prevents us from using data obtained after 1918. A suitable coherent run of data for a number of replicate plots comes from the 1890–1918 period. Climatic data is available for Columbia from the U.S. National Oceanic and Atmospheric Administration, currently from 1890. The average climate conditions at the Sanborn field have been very stable over the period 1895–1998, without any detectable trends in the temperature and precipitation. The average annual surface temperature in 1895–1998 was 13°C, with the maximum and minimum monthly mean temperatures of 26°C in July and about –2°C in January. Mean annual precipitation was 973 mm, and potential evapotranspiration, 790 mm (Hu and Buyanovsky 2003).

Chernozems and podzolic chernozems are widespread in the CTU area. Chernozems in the USA are classified within the Mollisol group (Fanning and Fanning 1989) and Sanborn lies at the south-eastern edge of the zone. Currently the detailed classification of the soil is an udollic ochraqualf, the mollic properties of the thin loess deposit being modified by the underlying glacial till; the top layer of the soil profile contains 2.5–2.9% organic matter (Hu and Buyanovsky 2003).

The yield (here denoted *Y*, in tonne/ha/year, with $Y_u$ obtained without any fertiliser and $Y_m$ obtained from manured plots) is known for each replicate plot between 1890 and 1918 (Miller and Hudelson 1921). Measurements of total rainfall between January and May are available at the experiment location (denoted *R*, in mm/year), and the time since the start of cultivation is known for each plot (denoted *D*, in years). These data are analysed below separately for plots with and without manure fertiliser applied, and where the wheat was grown every year, biennially or in rotation with other species. Data on the air and soil temperature at the Sanborn experiment site are also available. However, we do not use the temperature data in our analysis since the rainfall and temperature are not independent variables; on average, lower rainfall implies higher temperature. We use the rainfall data for the January–May period when the growth of the wheat is most critically affected by either drought in the early summer period (Arnon 1972) or by excess water leaching nitrogen from the soil (Hall 1905).

## 4.2 Variability and systematic trends of wheat yield

The data from the Sanborn experiment come from seven replicate plots of land, five treated with manure and two unmanured, with wheat grown annually.

### 4.2.1 Yield without fertilisers

For unmanured wheat grown every year at Sanborn, the average yield is 0.9 tonne/ha/year with a standard deviation of 0.7 tonne/ha/year (the coefficient of variation of 80%). The yield variability is very large, with a peak frequency at about 0.6 tonne/ha/year and a long positive tail (that is a few years gave exceptionally high yields). There are significant negative correlations between the wheat yield $Y_u$ and both rainfall from January to May *R* and the duration of cultivation *D*. The experimental data are shown with open circles in Figures 1a and 1b.

Assuming that the soil fertility is depleted by the same fraction each year, it might be expected that the dependence of the yield on time, and perhaps rainfall, is exponential. However, because of the large data scatter and relatively narrow ranges of the independent variables, it is more reasonable to adopt the simplest linear dependence of the yield at the unmanured plots, $Y_u$, on the January–May rainfall, *R*, and the cultivation duration, *D*,

$$Y_u = A + BR + CD, \tag{1}$$

with the constants *A*, *B* and *C* to be determined by fitting this dependence to the data. It is difficult to justify a more complex model given the data available. A least squares fit to the data from unmanured fields, shown in Figures 1a and 1b has the form

$$\frac{Y_u}{1\,\text{kg/ha/year}} = (2500 \pm 570) - (2.9 \pm 0.14)\left(\frac{R}{1\,\text{mm/year}}\right) - (40 \pm 14)\left(\frac{D}{1\,\text{yr}}\right), \tag{2}$$

where the uncertainties represent one standard deviation obtained from the scatter of the data points around the fit. The values of $Y_u$ obtained from this fit for the corresponding values of *R* and *D* are shown in the figures with filled circles to appreciate the quality of the fit.

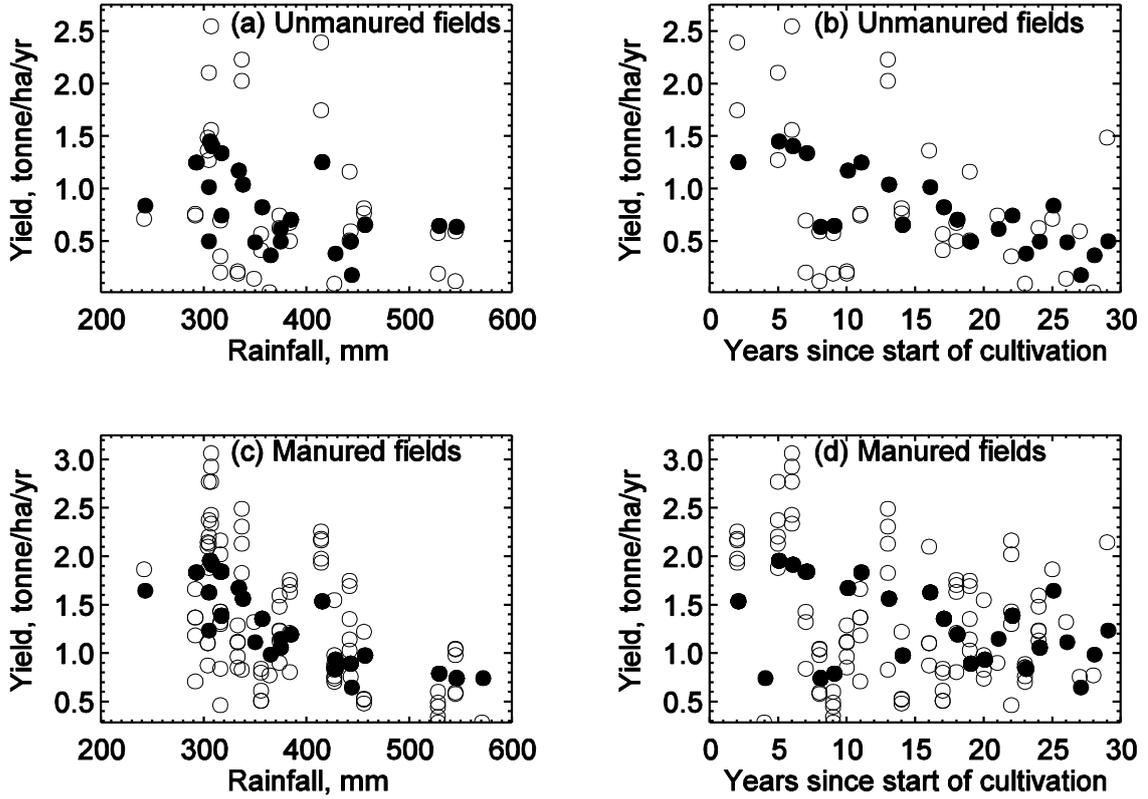

Figure 1. The dependence of wheat yield, $Y_u$ for unmanured (a, b) and $Y_m$ for manured (c, d) Sanborn plots, as a function of the annual rainfall $R$ (a, c) and the duration of continuous cultivation $D$ (b, d). Open circles show the experimental Sanborn data whereas filled circles represent fitted values calculated using Eqs. (2) and (3) for the corresponding values of $R$ or $D$ as appropriate. One outlying data point with $R = 142$ mm is not shown in Panels (a) and (c) and not included in the fit.

Figure 1a shows the yield (both observed and fitted) versus the January–May rainfall, and Figure 1b presents the variation of the yield with time after initial planting. The yield decreases, on average, with both $R$ and $D$. Rainfall over the period January to May averages to 400 mm with a standard deviation of 114 mm; it is clear that the higher rainfalls are not beneficial; the same effect was found at the Broadbalk experiment in England (Hall, 1905), where yield was reduced in wetter seasons. The rainfall of about 300 mm is nearly optimal for the crops as more rainfall just removes nutrients from soil in winter. The rainfall at Sanborn was less than 292 mm in only two years (256 mm in 1901 and 142 mm in 1914) that showed significantly reduced yields. However, the data available are not sufficient to identify such a non-monotonic dependence of $Y_u$ on $R$. Conservatively, the fits presented here should only be applied for $R \geq 300$ mm.

The reduction in yield with the cultivation time on these unmanured plots is not unexpected, and a similar reduction is clearly noted for the Urrbrae wheat experiment in Australia (Grace and Oades 1994).

This analysis relates to all unmanured replicate plots combined. To ensure that the trends are consistent across individual plots, we repeated the analysis for the two individual unmanured replicate plots. The fit to the data from Plot 2 has much larger errors, but the trends with rainfall and time remain. We show the fit coefficients and their errors for the individual plots and the summary results in Table 3.

Table 3. Fits to the yield data for individual plots and for the overall yield from all the plots: the fitted parameters *A*, *B* and *C* of Eq. (1), together with their respective standard deviations $\sigma_A$, $\sigma_B$ and $\sigma_C$. Note that the unit chosen is based on kilogram rather than tonne as used elsewhere in the text. The value of $\mathcal{R}^2$, given as a percentage, indicates the fraction of the variation in the data accounted for by the fit (higher values of $\mathcal{R}^2$ indicate a better fit, with the maximum of 100).

|  | *A* | $\sigma_A$ | *B* | $\sigma_B$ | *C* | $\sigma_C$ | $\mathcal{R}^2$ |
|---|---|---|---|---|---|---|---|
|  | kg/ha/year | | kg/ha/year/mm | | kg/ha/year² | | % |
| **Unmanured** | | | | | | | |
| **Both Plots** | 2500 | 570 | −2.9 | 1.4 | −40 | 14 | 27 |
| **Plot 1** | 2100 | 670 | −2.8 | 1.6 | −24 | 14 | 22 |
| **Plot 2** | 3000 | 1000 | −2.9 | 2.4 | −63 | 39 | 27 |
| **Manured** | | | | | | | |
| **All Plots** | 3500 | 310 | −4.7 | 0.7 | −30 | 8 | 37 |
| **Plot 1** | 2900 | 740 | −2.2 | 1.8 | −46 | 21 | 33 |
| **Plot 2** | 4300 | 820 | −6.0 | 2.0 | −38 | 23 | 46 |
| **Plot 3** | 3600 | 750 | −5.0 | 1.8 | −24 | 21 | 39 |
| **Plot 4** | 3900 | 650 | −4.9 | 1.6 | −48 | 18 | 53 |
| **Plot 5** | 3100 | 590 | −4.2 | 1.3 | −12 | 13 | 35 |

### 4.2.2 The effect of manure fertilisation

We used data from five manure-fertilised plots with wheat grown every year. These received 15 tonne/ha/year of farmyard manure but were otherwise identical to the unmanured plots. The average yield of all unmanured plots is 1.34 tonne/ha/year with a standard deviation of 0.7 tonne/ha/year (coefficient of variation of about 50%). The variability between the plots and years is much smaller than that in the unmanured plots. The yield $Y_\mathrm{m}$ is significantly correlated with the January–May rainfall *R* and time span *D* since the beginning of cultivation:

$$\frac{Y_\mathrm{m}}{1\,\mathrm{kg/ha/year}} = (3500 \pm 300) - (4.7 \pm 0.7)\left(\frac{R}{1\,\mathrm{mm/year}}\right) - (30 \pm 8)\left(\frac{D}{1\,\mathrm{yr}}\right) \quad (3)$$

Figure 1c and 1d show the yield data for manured plots (open circles) together with this fit (filled circles). Panel (c) shows the yields (both observed and fitted) versus rainfall, Panel (d) presents the yield dependence on time. The rate of decrease in yield with time is smaller than for the unmanured plots, while that with increased rain is larger. It is not surprising that the decrease with time is slower than that for the unmanured plots, as the manure supplied a large part of the nutrients removed in the harvested crop. The stronger decrease with rainfall can occur because more nutrients are leached from the soil in the wetter years, or because the thicker crop lodged (was knocked down) more severely by intense rain. The yield is, most frequently, higher than for the unmanured plots, and there is a long positive tail of infrequent very high yields. Again, each plot was analysed individually as well as collectively with all other manured plots. Similar trends are present at all replicate plots, as shown in Table 3.

As mentioned above, we also tried fits with exponential dependencies on rainfall and time span, but this did not improve the statistical quality of the results. The time span available (only around 25 years) is too short to make it practical to distinguish exponential and linear dependences. We note, however, that it is probable that the decline in productivity is exponential in the long term (i.e., there is a constant annual fractional decrease in yield).

For completeness, we also fitted a constant to the data, to test the hypothesis that the yield is independent of the rainfall and time; the resulting fits were significantly worse than the linear fits given above, confirming that the systematic trends revealed are meaningful.

Table 4. The cross-correlation matrix between the yield ($Y_u$ and $Y_m$ for unmanured and manured plots, respectively), rainfall ($R$) and time since the beginning of cultivation ($D$), denoted in the text $C_{ij}$ with $i, j = Y, R, D$. Larger correlation coefficients (by magnitude) indicate stronger statistical dependence between the corresponding variables; negative values indicate an anti-correlation (i.e., one variable decreases as the other increases).

| | Unmanured Plots | | | | Manured Plots | | |
|---|---|---|---|---|---|---|---|
| | $Y_u$ | $R$ | $D$ | | $Y_m$ | $R$ | $D$ |
| $Y_u$ | 1 | | | $Y_m$ | 1 | | |
| $R$ | −0.26 | 1 | | $R$ | −0.42 | 1 | |
| $D$ | −0.31 | −0.33 | 1 | $D$ | −0.18 | −0.30 | 1 |

To provide an additional measure of the yield sensitivity to the rainfall and time lag, we calculated the Pearson cross-correlation coefficient $C_{ij}$ between these variables. The cross-correlation coefficients given in Table 4 suggest that, in the case of unmanured plots, the yield is slightly more sensitive to time elapsed since the start of cultivation than to the rainfall: $|C_{YD}| > |C_{YR}|$, with $C_{YD} = -0.31$ and $C_{YR} = -0.26$. We note, however, that the difference is rather small and perhaps statistically insignificant. However, the opposite inequality applies to manured plots, where $|C_{YR}| = 0.42$ is more than a factor of two larger than $|C_{YD}| = 0.18$. Thus, the dependence on the rainfall dominates over the dependence on the time span in the variability and long-term trend of the yield from manured plots. The correlation between the rainfall and time span is similar for both manured and unmanured plots, $C_{RD} = -0.33$ and $-0.30$, respectively, which is a natural consequence of identical climate trends and the difference has no practical significance.

The relatively small values of $\mathcal{R}^2$ in Table 3 indicate that the yield can significantly depend on other variables apart from the rainfall and the time span. For example, our assumption that the temperature and rainfall are strongly negatively correlated, and thus are not independent variables, may be questionable. Hu and Buyanovsky (2003) note that, in the study area, higher temperatures often occur concurrently with increased rainfall. The relatively low values of the cross-correlations $C_{YR}$ and $C_{YD}$ in Table 4 are consistent with this suggestion. This question clearly deserves further analysis.

We also considered plots of biennial wheat crops, manured or unmanured, with clover as the intervening crop. There are fewer measurements available than for the monoculture wheat described above, and although the manured plots had a larger yield (1650 kg/ha/year as opposed to 1340 kg/ha/year at the unmanured plots) there is no qualitative change in the yield trends with either the passage of time or the amount of rain that fell.

The data summarised above are similar to those from other experiments, albeit in different climatic regions — Broadbalk in England (Hall 1905) and Urrbrae in the coastal belt of Australia (Grace and Oades 1994) — showing comparable response of the crops to the environment in such disparate areas, even if the trends may differ quantitatively. The very large variability of yield at Sanborn on the monocultural plots was explained by pest and disease attack and weeds (Miller and Hudelson 1921). The yield at Rothamsted farm in England (Hall, 1905) was less variable from year to year, probably because the impact of the outbreaks of pests and diseases was weaker in the cooler climate. Similarly to our results, Hu and Buyanovsky (2003) find that the *corn* yield at Sanborn was higher in years with lower rainfall in April and higher rainfall in May–August. They conclude that the corn yield is favoured by warmer and dryer spring months (April and May) and wetter and cooler July and August. These authors also find that "the average growing season climate gives little indication of climate effect on corn yield", and the yield variations are mainly controlled by monthly and shorter climate variations.

## 4.3 Adjustments to pre-modern agriculture

The Sanborn data have been obtained for relatively modern wheat varieties. [Unfortunately, Miller and Hudelson (1921), our main data source, do not identify the specific wheat varieties used in the experiments.] Even if the soil and climate conditions can be taken to be broadly similar to those of the CTU area, significant corrections are required to allow for the difference in the crop species and agricultural techniques. Nikolova and Pashkevich (2003) and Pashkevich and Videiko (2006) present and discuss evidence that the main cereal crops of the CTU farmers were hulled wheats, such as emmer (*Triticum dicoccum* Schrank), einkorn (*T. monococcum* L.) and spelt (*T. spelta* L.), as well as barley varieties (*Hordeum vulgare* and *Hordeum vulgare* var. *coeleste*).

Considering adjacent temporal and geographical domains, the cereal crop assemblages in early Neolithic cultures in Bulgaria (the second half of the sixth millennium BC) include naked and hulled barley (*Hordeum* sp.) and naked wheat (*T. aestivum* s.l./*durum*/*turgidum*), together with pulses, in addition to those cultivated by the LBK farmers: emmer (*T. diococcum*), einkorn (*T. monococcum*), as well as peas, lentils and flax (Kreuz et al. 2005). These authors note that barley and naked wheat were used in the broader area, including that of the Starčevo–Körös–Čris culture (eastern Hungary, Greece, former Yugoslavia, Romania and the Turkish Thrace). A review of other estimates of the wheat yields, including experimental, historical and ethnographic data can be found in Table 2.1 of Bogaard (2004b). Her data are generally consistent with our estimates, especially given the fact that they refer to naked wheat varieties and barley, whereas we focus on hulled wheats.

Pashkevich and Videiko (2006) suggests that the CTU farmers relied on spring crops and did not use winter crops. From the potential weed species recorded at the Neolithic sites (in particular, winter annuals versus summer annuals), Kreuz et al. (2005) conclude that both summer and winter crop growing was typical of the early Bulgarian Neolithic, whereas summer crop cultivation apparently dominated at the LBK sites. The Sanborn crops considered in Section 3 are winter crops. We note that winter crops have higher yields than spring varieties on the same land (by 25% or more – Percival 1974, p. 422) but, correspondingly, they deplete the soil fertility more than the summer crops. As a result, growing winter crops often requires crop rotation, which reduces the yield averaged over a sufficiently long period. There are certain disadvantages of winter crops as compared to summer ones: the fields need to be prepared for sowing in a rather short time, and winter crops are more sensitive to climate fluctuations. A certain balance of winter and summer wheats appears to be optimal.

Table 5. Yields of spring emmer and einkorn, and winter spelt, together with naked wheat yields grown under comparable conditions, under dryland cropping in south central Montana, U.S.A. in 1992–1994 (emmer and einkorn) and 1991–1994 (spelt) (after Stallknecht et al. 1996). The emmer, einkorn and spelt grain yields were estimated as 60% of the hulled grain when dehulled.

| Wheat variety | Mean grain yield | Yield range | Relative to naked wheat |
|---|---|---|---|
| Unit | kg/ha/year | kg/ha/year | % |
| Emmer [a] | 1990 | 1540–2550 [e] | 58 |
| Einkorn [a] | 2600 | 120–4160 [f] | 76 |
| Naked wheat [b] | 3417 | 2250–5370 [e] | 100 |
| Spelt [a,c] | 3040 | 2090–4240 [e] | 72 |
| Naked wheat [c,d] | 4233 | 3430–5910 [e] | 100 |

**Notes:** [a] Data for five highest-yielding selections. [b] 'Newana' hard red spring wheat. [c] Winter multi-year yields. [d] 'Tiber' hard red winter wheat. [e] Range of average yields over a set of plots. [f] Yield range for individual plots.

## 4.3.1 Correction for the wheat varieties

Stallknecht et al. (1996) provide data on the yield of selected crossings of emmer, einkorn and spelt grown at the Southern Agricultural Research Center, Huntley, Montana, U.S.A. in 1991–1994. These modern varieties were selected for their *high yield*, so the data, summarised in Table 5, should be used with great caution in the present context. The yields of einkorn, emmer and spelt are significantly lower than those of modern naked wheats grown under similar conditions; the data of Table 5 suggest that the yields of even the best selections of emmer and spelt are 60–75% of naked wheat yields. We also note the strong variability of the yields, shown in Column 3 of Table 5 in terms of the yield range. The range for einkorn is based on the data series for individual plots, and shows variations by about 100%, whereas the other entries show the range of the annual averages over a set of plots, thus showing less variability, at about 25% (if the individual plots have 100% variability, such a reduction could be achieved with 10–15 plots in each set).

Percival (1974, pp. 171 and 188) estimates the einkorn yield as 16–80 hectolitres per ha (about 1200–6000 kg/ha/year) depending on the soil quality (ranging from poor mountainous regions to good soils), whereas emmer yields vary from 25 to 50 bushels per acre (about 1700–3400 kg/ha/year). The largest einkorn yield given by Percival is significantly higher than that in Table 5, but the emmer yield is in a better agreement with Table 5. We stress again that the emmer, einkorn and spelt data in Table 5 and those of Percival (1974) are at the higher end of the range even for the modern plant varieties.

Jarman et al. (1982, p. 158) quote historical data on the average cereal yield of 800–1400 kg/ha/year in traditional agricultural systems in Romania and note its strong fluctuations from about 1400 kg/ha/year in 1913 to 540 kg/ha/year in 1914. Nikolova and Pashkevich (2003) quote the emmer yields for 1902 in south Ukraine at the level of 390–1140 kg/ha/year; with the median value (750 kg/ha/year) significantly smaller than that given in Table 5. Russell (1988, p. 111) suggests, for the early agriculture in the Near East and Africa, 500 kg/ha/year for the emmer and spelt yields, with a range of 400–3700 kg/ha/year. Gregg (1988, pp. 73–74) quotes the range of 757–1045 kg/ha/year for the late nineteenth century yields of winter and spring einkorn and emmer–spelt maslin in Germany, and adopts the larger value in her estimates for the LBK agriculture. The yields of autumn-sawn emmer in the Butser Ancient Farm experiment averaged over 15 consecutive seasons at about 2080 kg.ha/year, grown without using manure on a field every second year with a bean crop in between (Reynolds 1992). The author notes a rather high yield, "significantly higher than any expectations", attributable to "the soil, the climate and good management".

Karagöz (1996) provides data on the yield of einkorn and emmer in Turkey in 1948–1993. Although the data are only given for the two species combined, the author notes that emmer was planted on much larger areas than einkorn. According to this author, the yield varied from 814 to 1391 kg/ha/year, with the mean and standard deviation of 1110 ± 200 kg/ha/year. This variation was not uniform in time: the yield did not change much in 1948–1968 when it was 930 ± 100 kg/ha/year, but exceeded 1231 kg/ha/year thereafter.

Karagöz (1996) also reports an agricultural experiment in northern Turkey, with very limited use of fertilisers and herbicide. Naked wheat was grown on 1280 ha, and emmer and barley on 542 and 456 ha, respectively, in sloping, marginal forest areas. The average yields of naked wheat, barley and emmer in this experiment were 847, 711 and 618 kg/ha/year, respectively. The modern annual average rainfall in the area is 567 mm, and the average annual temperature is 10.4ºC; the soil cover is predominantly the Brown Forest Soil.

In another experiment (Castagna et al. 1996), einkorn gross yield (i.e., that of hulled grain) varied broadly between 840 and 4570 kg/ha/year (with a typical value of 2840 kg/ha/year), with the net yield estimated as 77% of the gross value on average. The maximum gross grain yield was obtained with a seeding rate of 72 kg/ha/year (300 kernels/m$^2$/year). The yield of two bread wheat cultivars (*T. aestivum*) grown as controls averaged at 7030 kg/ha/year.

Considering also the other extreme, we note that the yield of *wild* einkorn and emmer can reach 500–1000 kg/ha/year (see Araus et al. 2007, and references therein). Araus et al. (2007) use the stable carbon isotope ratio $^{13}C/^{12}C$ in the fossil grains of naked wheat (*T. aestivum/durum*) recovered from early Neolithic sites to estimate the prehistoric grain yield. The total number of 54 grains from Tell Halula and Akarçay Tepe (8000–6100 BC, Middle Euphrates region) were used for this purpose. This method relies on the strong connection, observed in modern wheat crops, between both the total water inputs during grain filling and grain yield, on one side, and the (normalised) difference in $^{13}C/^{14}C$ between the grain kernels and atmospheric $CO_2$, on the other side (Araus et al. 2003). The atmospheric carbon isotope content of the time was obtained by the authors from the Antarctic ice-core records. Furthermore, ancient soil fertility and/or the occurrence of fallow can be estimated from the grain $^{15}N/^{14}N$ ratio. The estimated wheat yield is 1300–1700 kg/ha/year, comparable to or even higher than that of modern wheat varieties in this region grown without irrigation. This can be attributed to a favourably wetter Neolithic climate in the area or to planting in alluvial areas. Furthermore, high values of $^{15}N/^{14}N$ in the ancient grain suggest that it was grown on fertile soils, perhaps with manure application and/or the use of natural wet soils. Altogether, Araus et al. (2007) suggest that the yield of naked wheat in the early agriculture in the area studied could plausibly be as high as 1000 kg/ha/year (see also Araus et al. 2001).

Given the differences in agricultural technologies and especially the wheat varieties from the modern experimental farms, it is fair to assume that the yields of the CTU crops were significantly lower than those of the Sanborn data of presented above. The relation between the yields of naked and hulled wheats grown under similar conditions that follows from Table 5 suggests that the yield of einkorn, emmer and spelt can be adopted as 70% of the ancient naked wheat yield estimated by Araus et al. (2007), i.e., of order 700 kg/ha/year. Incidentally, this figure is close to the emmer yield in the early twentieth century Ukraine quoted above, and somewhat smaller than the lower-end yields of emmer and einkorn in modern agricultural experiments. Whenever required, we shall allow for this correction by multiplying the yield of Eqs. (2) and (3) with a factor $\varepsilon$ chosen as to adjust the average yield at unmanured Sanborn plots, 900 kg/ha/year, to about 700 kg/ha/year. This yields

$$\varepsilon \approx 0.8.$$

This appears to be a very conservative estimate of the correction for the yield of cereals in the Neolithic: the yield could be noticeable larger, i.e., $\varepsilon$ can be larger.

Table 6. Fit parameters and their standard deviations for Eq. (4), based on the wheat yields at Sanborn given in Table 3 and Eqs. (2) and (3), with and without manure fertilization.

| Parameter | $Y_0$ | $\sigma_Y$ | $R_0$ | $\sigma_R$ | $D_0$ | $\sigma_D$ |
|---|---|---|---|---|---|---|
| Unit | kg/ha/year | | mm | | year | |
| **Unmanured plots** | 2539 | 571 | 879 | 457 | 64 | 26 |
| **Manured plots** | 3540 | 306 | 758 | 134 | 117 | 32 |

### 4.3.2 Adjusted yield trends with the rainfall and the cultivation time

We shall be using the trends given in Eqs. (2) and (3), being aware of the tentative nature of these results. Rewriting these equations in a more convenient form, we shall be using fits of the following form for $Y_u$ and $Y_m$:

$$Y = \varepsilon Y_0 \left(1 - \frac{R}{R_0} - \frac{D}{D_0}\right), \tag{4}$$

where $\varepsilon$ is the correction factor suggested above, and the fitted values of $R_0$ and $D_0$ are given in Table 6, as obtained from the fits for all unmanured and manured plots in Table 3. Here $R_0$ and $D_0$ have an intuitively clear meaning of the nominal values of the rainfall and the time span, respectively, required to reduce the yield to zero if only one of the two parameters varies while the other is fixed at zero.

For comparison, Percival (1974, p. 420) provides an approximation to the dependence of the average wheat yield in Britain in 1884–1904 on the total rainfall in October–December: yield per acre equals 39.5 bushels minus 5/4 of the rainfall expressed in inches, which translates into $Y_0 = 2660$ kg/ha/year/ and $R_0 = 800$ mm, figures rather similar to those in Table 5.

Jarman et al. (1982, p. 141) refer to the Rothamsted Broadbalk continuous wheat experiment (where the soil is a chalk-rich loam) suggesting "that, even without manure or fertiliser, average yields of grain showed only a very gradual decline over 60 years". The data shown in their Fig. 52 exhibit a decrease in the yield from 9 to 5–6 cwt/acre/year (1130 to 630–750 kg/ha/year) in about 20 years, followed by a variation between the latter value and 7 cwt/acre/year. Our fits for unmanured Sanborn plots give a decrease in yield by 50% in about 30 years, in a reasonable agreement with the initial decrease in the Rothamsted Broadbalk experiment. However, Loomis (1978) argues that, for a lower wheat yield of about 1000 kg/ha/year, nitrogen removed by the wheat crop (20 kg N/ha annually) is replaced during a crop–fallow cycle by dust, rain and birds (8–12 kg N/ha/year), by the seed (1 kg N/ha/year for the yield/seed ratio of 10 to 1), and by leguminous weeds (2–10 kg N/ha/year) and manuring. As a result, the nitrogen budget can be balanced and remain in equilibrium even without manuring (see also Gregg, 1988, p. 65). Loomis refers to existing cropping systems in Asia that have maintained such equilibria through thousands of years and notes that plots in Rothamsted experiment generally stabilized at a low yields of 1000–2000 kg/ha/year without manuring. The Sanborn data series is too short to assess this suggestion: the yield of unmanured plots in Figure 1b do not show any signs of reaching any equilibrium value in 30 years of cropping, whereas the manured plots of Figure 1d may have reached it in 15–20 years.

We stress that the values of $R_0$ and $D_0$ have been obtained from our fits to the Sanborn data, and we are unable to apply any corrections to make them better applicable to the opremodern CTU agriculture, even if such a correction can be reasonably introduced for $Y_0$. Admittedly, this is not satisfactory, but we are not aware of any data or arguments which would help to resolve the problem. On the other hand, the trends with time and rainfall can be less sensitive to the wheat variety than the yield.

### 4.3.3 On the use of the manure fertiliser

Having noted the strong variability of the yield, evident from Figure 1 (see also Nikolova and Pashkevich, 2003), we suggest that the Neolithic farmer would experience a boom and bust production system which could be mitigated to some extent by the use of manure. There is ample evidence for the use of manure as a fertiliser from the early stages of farming (Wilkinson 1982; Bogaard et al. 2007, 2013; Vaiglova et al. 2014). However, as there would be (at least initially) a large area of virgin land available that was relatively easy to clear for the fields, the extra work of collecting and using manure could have been avoided by the use of fresh fertile soil in new fields. In addition, the possibility of collecting manure in useful quantities depends on how the livestock is kept, and often requires that the cattle be brought to barns every night; this may or may not have been the practice in the CTU settlements. However, as the manure helped to reduce yield variability from year to year, this could make its use much more advantageous. In the Sanborn data, yields smaller than 400 kg/ha/year occurred on fewer than 8% of occasions under manure, but on 27% of occasions on the unmanured plots. It can be argued that large, relatively short-term, negative fluctuations in the productivity, rather than its general low level, can lead to catastrophic consequences and affect the survival and subsistence strategy and patterns of the population (Feynman and Ruzmaikin 2007; Abbo et al. 2010). The fact that manuring stabilises the yield under variable environmental conditions could make the use of the fertiliser an especially attractive option for the Neolithic and CTU farmers. We estimate below the maximum fraction of the crop area that could be manured given the herd composition of the CTU farmers,

## 5. The diet of CTU farmers

### 5.1 Cereals

Following the results of Section 4, we adopt $Y = 700$ kg/ha/year as a nominal yield of hulled wheats, but consider plausible the range of 700–1200 kg/ha/year; even higher yields may be appropriate, especially for later CTU stages. Emmer seeding rates are 76 kg/ha in low-rainfall regions and 100 kg/ha in high-rainfall areas; 67–100 kg/ha is the seeding rate of spelt on dryland (Stallnecht et al. 1996). Einkorn seeding rate is similarly about 72 kg/ha/year (Castagna et al. 1996). These estimates agree with the general figure of about 10% or more of a harvested grain to be used as seed crop (e.g., Hillman and Davies 1990, p. 178). We adopt the seeding rate of 12% in our calculations. For comparison, White (1963) suggests, based on documentary evidence (Varro), the wheat yield in Roman Etruria was between ten- and fifteen-fold. Assuming that further 25% of the grain is lost to pests (Hall, 1905), about 440 kg/ha/year remains available for consumption.

The World Health Organisation proper nutrition recommendation of 2200–3000 kcal/person/day translates into about 900–1200 kcal/person/day from each of domestic animal products and cereals, assuming that each contributes 40% of the calorific diet content. Using the calorific value of the spelt grain of about 3150 kcal/kg (Ranhotra et al. 1996), the required amount of cereals is 100–140 kg/person/year. With the grain available as food of 440 kg/ha/year, this implies the required crops area of about 0.2–0.3 ha/person. Although emmer and einkorn dominate over spelt at the CTU sites, the calorific content of their grain, 3567 kcal/kg for einkorn (Harlan 1967, p. 198), does not differ much from that of spelt; we conservatively adopt the lower figure.

Palaeoeconomy estimates often neglect the contribution of domestic and wild animal products to the diet and assume (explicitly or implicitly) that cereals are the only component of the Neolithic diet. Using the above figures, 250–350 kg/person/year of cereals would be required as the sole source of calories, which would need the area of 0.4–0.5 ha/person to produce if any losses are neglected (as is done equally often). This figure is similar to many earlier results, which we believe to be overestimates.

Table 7. Animal bone assemblages from Trypillia sites: the minimum numbers of individuals (MNI) at the sites specified in Column 1 (after Appendices 2–5 of Tsalkin 1970) and mean and relative numbers for each Trypillia stage (bold). Data are given here only for the animals suitable as a food resource and occurring in significant numbers. The relative mean MNI values and their standard deviations are given separately for the domestic and wild animals.

|  | Cattle | Sheep/goat | Pig | Horse | Total domestic | Red deer | Roe deer | Wild boar | Total wild |
|---|---|---|---|---|---|---|---|---|---|
| *Early Trypillia (Stage A)* | | | | | | | | | |
| Sabatinovka 2 | 22 | 11 | 10 | 9 |  | 8 | 4 | 3 |  |
| Luka-Vrublevetskaya | 42 | 38 | 93 | 4 |  | 57 | 31 | 33 |  |
| Bernovo-Luka | 23 | 9 | 11 | 2 |  | 25 | 17 | 17 |  |
| Lenkovtsy | 30 | 10 | 19 | 5 |  | 25 | 9 | 9 |  |
| Soloncheny I | 17 | 14 | 19 | 3 |  | 20 | 7 | 15 |  |
| Galerkany | 5 | 4 | 1 | 3 |  | 6 | 2 | 6 |  |
| Karbuna | 11 | 7 | 6 | 4 |  | 2 | 2 | 2 |  |
| *Mean MNI* | 21 | 13 | 23 | 4 | 62 | 20 | 10 | 12 | 43 |
| *Relative mean MNI* [a] | 0.35 | 0.22 | 0.37 | 0.07 | 1.00 | 0.48 | 0.24 | 0.28 | 1.00 |
| *Standard deviation of the relative mean MNI* [a] | 0.09 | 0.05 | 0.14 | 0.08 |  | 0.08 | 0.07 | 0.086 |  |
| *Middle Trypillia (Stages BI, BI-II and BII)* | | | | | | | | | |
| Sabatinovka 1 | 30 | 14 | 14 | 9 |  | 10 | 4 | 8 |  |
| Berezovskaya GES | 12 | 6 | 6 | 3 |  | 20 | 3 | 8 |  |
| Soloncheny II | 39 | 14 | 26 | 6 |  | 34 | 7 | 20 |  |
| Khalepje | 11 | 17 | 8 | 2 |  | 2 | – | 1 |  |
| Kolomijshchina II | 8 | 5 | 3 | 2 |  | 1 | – | 2 |  |
| Vladimirovka | 36 | 30 | 25 | 5 |  | 11 | 3 | 1 |  |
| Polivanov Yar | 33 | 39 | 92 | 3 |  | 24 | 14 | 16 |  |
| *Mean MNI* | 24 | 18 | 25 | 4 | 71 | 15 | 6 | 8 | 29 |
| *Relative mean MNI* [a] | 0.34 | 0.25 | 0.35 | 0.06 | 1.00 | 0.51 | 0.22 | 0.28 | 1.00 |
| *Standard deviation of the relative mean MNI* [a] | 0.10 | 0.09 | 0.13 | 0.04 |  | 0.14 | 0.07 | 0.18 |  |
| *Late Trypillia (Stages CI and CII)* | | | | | | | | | |
| Podgortsy 2 | 11 | 6 | 16 | 6 |  | 1 | – | 1 |  |
| Syrtsy | 5 | 12 | 1 | 1 |  | – | 3 | – |  |
| Koshilovtsy | 5 | 2 | 3 | 1 |  | – | 1 | – |  |
| Sukhostav | 3 | 1 | 2 | 1 |  | – | – | – |  |
| Usatovo | 266 | 438 | 25 | 163 |  | 16 | 3 | 6 |  |
| Starye Bezradichi | 2 | 3 | 2 | 1 |  | 2 | 1 | 1 |  |
| Kunisovtsy | 5 | 3 | 3 | – |  | 2 | – | 1 |  |
| Andreevka | 3 | 4 | 2 | 1 |  | 1 | – | – |  |
| Sandraki | 4 | 5 | 3 | 3 |  | 3 | 3 | 5 |  |
| Stena | 13 | 9 | 14 | 9 |  | 6 | 4 | 5 |  |
| Gorodsk | 14 | 8 | 14 | 7 |  | 4 | 6 | 3 |  |
| Troyanov | 13 | 7 | 6 | 5 |  | 4 | 2 | 4 |  |
| Pavoloch' | 6 | 6 | 3 | 1 |  | 3 | 1 | 1 |  |
| Kolomijshchina I | 12 | 8 | 7 | 3 |  | – | – | – |  |
| Podgortsy I | 12 | 2 | 2 | 2 |  | 5 | 1 | 3 |  |
| *Mean MNI* [a,b] | 8 | 5 | 6 | 3 | 22 | 3 | 2 | 3 | 8 |
| *Relative mean MNI* [a,b] | 0.35 | 0.25 | 0.25 | 0.14 | 1.00 | 0.38 | 0.30 | 0.32 | 1.00 |
| *Standard deviation of the relative mean MNI* [a,b] | 0.11 | 0.14 | 0.09 | 0.06 |  | 0.21 | 0.35 | 0.10 |  |
| *Grand total* | | | | | | | | | |
| Relative mean MNI [a,b] | 0.35 | 0.24 | 0.33 | 0.08 | 1.00 | 0.47 | 0.24 | 0.29 | 1.00 |
| Standard deviation of the relative mean MNI [a,b] | 0.10 | 0.11 | 0.11 | 0.06 |  | 0.21 | 0.25 | 0.15 |  |

**(a)** Given separately for domestic and wild animals. **(b)** Excluding the Usatovo data

## 5.2 Domestic animal products

To estimate the size of cattle and caprine herds required to satisfy the nutrition needs of the Neolithic and Bronze Age farmer, we assume that the animals were kept for both meat, milk and dairy products (and perhaps blood). However, wildlife resources are another source of meat, and there is sufficient archaeological evidence, similar to that given in Table 7, to assume that wild animal meat was also an important source of nutrition. As discussed above, Ogrinc and Budja (2005) suggest that about 20% of the diet at the Ajdovska Jama site was provided by the meat of wild animals.

Zhuravlev (1990, p. 137) analysed the animal bone assemblage of Maydanetske, one of the largest CTU sites known (Trylillia CI, Cherkassy Region, central Ukraine) to estimate the fraction of domestic animals as 85% by head, comprising 35% of cattle (*Bos taurus L.*), 27% of sheep (*Ovis aries L.*) and goats (*Capra hircus L.*), 28% of pigs (*Sus domestica Gray*) and 5% of horses (*Equus caballus L.*); this appears to be a typical picture for both early and late Trypillia settlements in the Ukraine. These figures are encouragingly similar to those of Tsalkin (1970) presented in Table 7. A very detailed and extensive overview of the CTU bone assemblages, their biometric characteristics and local variations can be found in Zhuravlev (2008) and Videiko et al. (2004, Vol. 1, pp. 152–198). These authors note a relatively large fraction of cattle in the apparent herd structure and suggest, from the osteometric data, that bulls, oxen and horses were used as draught animals.

There are several clear trends in the bone assemblages presented in Table 7. The ratio of domestic to wild animals (by MNI, the minimum number of individuals) increases from 1.4 in the Early Trypillia to 2.4 in the middle period and to 2.8 in the late stage. The composition of the domestic livestock apparently remains stable within errors, apart from the increase in the relative frequency of the horse MNI from small quantities in the Early and Middle stages to $0.14 \pm 0.06$ in the Late Trypillia. The faunal remains at Usatovo (Late Trypillia) are clearly exceptional (e.g., Zhuravlev 2008) and are excluded from the averages presented in the table.

For the herd/flock composition, we adopt the relative mean MNI numbers from the bottom of Table 7, $a_c = 0.35$ of cattle, $a_s = 0.24$ of caprines, $a_p = 0.33$ of pigs and $a_h = 0.08$ of horses in terms of the relative numbers by head. The energy content of the meat from the domestic animal species can be found in Gregg (1988, p 152) and Jarman et al. (1982). The average culling rate in the modern UK cattle herds is 25% (AHDB 2012); our nominal figure of the cattle herd fraction culled annually is $k_c = 0.2$; the culling rate of caprines, $k_s$, is assumed to 0.2 too. Since pigs are not kept for milk, their culling rate $k_p$ can be higher; but we adopt $k_p = 0.2$.

Following White (1953), we assume that a half of the live weight of both cattle and caprines represents usable meat; the figure for pigs is 0.7. The live weight of cattle and caprines adopted are 200 kg/head and 50 kg/head, respectively. Neolithic pigs were significantly smaller than either wild or modern ones. This difference, noted from the CTU bone assemblages by Tsalkin (1970, p. 179) and Zhurvalev (2008, p. 17) is interpreted as evidence that the pigs were isolated from their wild relatives using fences or pens. Following Gregg (1988, p. 118), we adopt 30 kg/head for a pig's live weight.

Bökönyi (1971) suggests that, in the Middle Neolithic, cows could provide only little surplus milk after the calf had been fed. This would of course depend on the feeding of the cow, and the size, vigour and the weaning age of the calf. However, dairy foods appear to be used in the Neolithic (Craig 2002; Copley et al. 2003; Craig et al. 2005; Spangenber et al. 2006; Evershed et al. 2008), and the importance of dairy farming apparently increased qualitatively in the Bronze Age (Sherratt 1983, 2010; Greenfield 2005; Brochier 2013). Milk was valued to the point that calves seem to have been weaned early during the Neolithic (Balasse and Tresset 2002). Composition of the milk is affected by the diet of the animal (Boland 2003), with those fed on grass without a concentrate feed having a lower yield, more butterfat and similar protein content. The breed and species also have a strong effect on milk composition (Crawford 1990), with modern breeds such as the Holstein having lower butterfat content.

It is difficult to estimate the milk yield in the CTU or any other prehistoric farming system. To start at the lower end of the modern productivity, we note that, in modern subhumid Nigeria, milk yield from 'traditional' cattle is 280.7 litres per annum of which 111.5 litres is a surplus to the calf's requirement (Otchere 1986). A figure of 0.59 litre/day (or about 215 litre/year) surplus for the Zebu cattle in Tanzania was reported by Kavana et al. (2006). 'Indigenous' cattle in Ethiopia on smallholdings produce a total of 1.5–3.6 litre/day with the average lactation length of 232 days (Abraha et al. 2009), as compared to 1.6–2.4 litre/day for 'indigenous' stock in Zimbabwe (Masama et al. 2003). It is notable that, in some of the above cases where the milk yield is very low, the cattle is kept mostly for prestige and other similar non-economic reasons. It is hard to find suitable European data since even in the less developed areas such as Moldova, the 'traditional' breeds produce nearly 10 times the above yield (Moldova 2004), and even the worst producer (in a survey of, predominantly, smallholders with less than 3 cows) was producing 1400 litre/year or more in 2001 and 2003 (Dumitrasko et al. 2006). Todorova (1978) suggests that a Neolithic cow produced some 600–700 litres of milk annually. Gregg (1988, p. 106) adopts a cow's milk yield of 1.78 litre/day, which leads to about 360 litre/year/head for a lactation length of 200 days. As a nominal figure, we adopt the surplus cow milk yield of $y_c = 400$ litre/head/year but consider a range of 0–2000 litre/head/year. For comparison, modern European cow breeds typically produce 10,000 litre/head/year of milk.

For the milk yield of sheep and goats, we adopt values at the lower end of the modern range. For a 12-week annual lactation period and hand-milking, non-dairy goats and sheep produce in Malawi 61 and 34 kg/head/year of milk, respectively (Banda et al. 1992). Gregg (1988, p. 118) quotes 170–680 kg/head/year for sheep and 340–1417 kg/head/year for goats (as they have a longer lactation period). We prefer to use the conservative lower estimates, and the nominal figure used in our calculations is a rounded mean of the figures of Banda et al., $y_s = 50$ litre/head/year. Since caprines represent a relatively small fraction of the livestock, this choice does not greatly affect our results.

Estimates of the cattle grazing area range from 1 ha/head/month in deciduous forests to 1.5 ha/head on pasture (Gregg 1988, pp. 106–107). Jarman et al. (1982, p. 108) adopt the grazing area required for cattle of about $A_c = 10$ ha/head but note that it can be as low as 0.3–0.5 ha/head on seasonally and permanently flooded pasture. Gregg (1988, p. 123) suggests that the grazing area required for the herd should be doubled to allow for at least one-year recovery of the grazing land. Glass (1991, p. 28) quotes a number of estimates of the forest pasture area ranging from 0.8 to 8 ha/head. We adopt $A_c = A_h = 10$ ha/head as the nominal figure for both cattle and horses; detailed knowledge of the landscape around specific sites would be required to refine this estimate. Caprines' needs in grazing are about ten times smaller than those of cattle. When kept in large herds and under extensive grazing systems, sheep and goat need about $A_s = 0.5$ ha/head of grazing area (Coop 1986); this is the figure we adopt. However, the grazing characteristics of cattle, sheep and goats are complementary, as cattle and sheep relish grasses and herbs, respectively, whereas goats prefer weeds and woody vegetation not used by the other animals (Coop 1986; Gregg 1988, p. 123). We neglect any pasture area for the pigs as they can graze in woodlands and/or near the rural settlements; to some extent, this also applies to goats.

Fodder for four winter months is another requirement of livestock imposing constraints on both the exploitation area and the labour costs. Apart from meadow hay, cereal straw and leaves of certain trees such as elm (Rasmussen 1990), elder, ash and acacia provide good fodder. Modern grass–legume pastures can yield up to 5–20 tonne/ha/year of dry hay (Coop 1986); mature cows consume about 400 kg/head/month of hay and sheep/goat require about ten times less food (Gregg 1988, pp. 108 and 118). Gregg adopts the yield of a natural meadow on low-lying damp soils to be 1470 kg/ha/year. We follow this author to assume that about $M_c = M_h = 0.5$ ha/head of hay meadow is required to produce winter fodder for cattle and horses, and $M_s = 0.02$ ha/head for sheep/goats (Gregg 1988, pp. 110, 120 and 121). Since not only natural or cultivated meadows but also forests are a source of leafy fodder, we assume that only half of the fodder is hay and cereal straw. We include the area required to produce hay into the calculations of the exploitation area of a settlement in Section 7, and the time to cut grass on them in the labour costs and labour return in Section 8.

### 5.3 Wild animal products

The faunal remains found at CTU sites indicate that hunting was a significant source of food, especially at the early CTU stages. The ratio of wild to domestic MNI in Table 7 decreases from about 0.7 in the Early Trypillia to 0.4 at later stages. A more recent analysis of Zhuravlev 2008) shows a lower fraction of wild animals, of order 0.2. We adopt this figure in our calculations. The composition of the hunting trophy given in Table 8 is taken according to the relative mean MNI in the bone assemblages: 0.48 of red deer (*Cervus elaphus L.*), 0.24 of roe deer (*Capreolus capreolus L.*) and 0.29 of wild boar (*Sus scrofa ferus L.*) by head. The calorific value of the meat is taken from Jarman et al. (1982, p. 83).

## 6. Land use and the local carrying capacity

In this section we estimate the land area required for a farming population to subsist in a given environment, with a given subsistence strategy and agricultural technology, and hence the maximum number of people per unit area. We call this the *subsistence carrying capacity, $K_s$*, as opposed to an economic behaviour aimed at creating a surplus product for exchange or trade. The starting point for such a calculation are the human dietary requirements.

Any estimate of the carrying capacity of a landscape strongly depends on the subsistence strategy and on the land use. Ethnographic evidence presented by Jarman et al. (1982, p. 30) suggests that land could be exploited within 1–11 km of a settlement. This radius is limited by the time required to travel to the field, with one hour as a reasonable maximum, and 1.5–2 hours as an undesirable upper limit (similarly to the commuting times of modern urban workers, as Jarman et al. note). The average outside limit of the cultivated land area is suggested as 5 km, with most land under cultivation within 1–2 km of the settlement. Higgs and Vita-Finzi (1972) suggest a radius of 5 km for the exploitation territory by a sedentary population (and 10 km, for sedentary or semi-sedentary people), and note that time spent on travel is more important than distance (see also Jarman et al. 1982, pp. 30–32). Tipping et al. (2009) carefully analysed and modelled pollen data from an early Neolithic site in north-east Scotland (a timber 'hall' at Warren Field), to conclude that land within a radius of at most 2.5 km was in use. Cereals were cultivated immediately around the 'hall', but no evidence of pasture for livestock has been recorded. Following Chisholm (1979, p. 72), Higgs and Vita-Finzi (1972), Jarman et al. (1982) and many other authors, we assume that the cultivated fields will tend to be located in a close proximity, within not more than about 5 km of the settlement, and preferably within 1–2 km. The livestock can be kept at larger distances: up to 5 km if walking to the pasture and returning to the farm daily, or 10 km if the animals are kept around a temporary camp.

The family size is another important parameter. Five to seven people is a reasonable estimate for the size of an extended farming family, of which 2–4 may be fit to work in the fields, the remaining being too young or too weak. We adopt six people in a family as a representative value. Although a few family members could be involved in the physically demanding work such as land tillage, many other production activities can be assigned to other family members. For example, a large proportion of the herding and care of the domestic animals can be assigned to children. Tillage with the ard or plough requires two people to work simultaneously, but guiding the draught animal(s) does not require much physical force. Likewise, reaping, threshing (especially using animals), winnowing and later preparation of grain could involve virtually the whole family. Therefore, our discussion of the labour costs and the seasonal time stress largely focuses on the land preparation for sowing, an activity that requires significant physical force and must be completed in a short and strongly limited time.

Gaydarska (2003) presents land use analysis of Maydanetske, a proto-urban site (Trypillia CI) that had an area of $A_0 = 210$ ha (Müller at al. 2014) and an estimated $N = 10{,}000$–$15{,}000$ inhabitants; sites that large are rare but not exceptional: the area of the nearby Tallianky is 350–400 ha. The giant settlements emerged at late Trypillia stages. Typical settlement areas at various Trypillia stages are given in Table 2. Houses in CTU settlements are often arranged along nearly elliptical contours closer to the settlement boundary (perhaps to provide easier access to the fields) with large open spaces in the central part of the settlement that could be used for horticulture. According to Gaydarska (2003), about 78% of the area within 7 km of Maydanetske is suitable for agriculture; thus, $\delta_u = 0.2$ appears to be an acceptable estimate of the fraction of unusable area in the central part of the CTU area in the Dnieper–South Buh interfluve. We further assume that a fraction $\delta_a = 0.35$ of the total land area is potentially arable; the rest can be used as grazing land. We further assume that part of the arable land lies fallow; the ratio of the fallow to cropped land areas is denoted $\delta_f$. The nominal value adopted is $\delta_f = 2$, that is any plot is cropped once in three years. As an example from another region, the LBK study area of Ebersbach and Schade (2004), Mörlener Bucht in Hesse north of Frankfurt am Main, has 82% of the area suitable for fields (loess soil), 11% are water meadows suitable for grazing and 7% are steep slopes suitable neither for fields nor for grazing.

To make our results properly robust and flexible, we first derive general algebraic expressions for the key variables involved in palaeoeconomy reconstructions before using specific values of the input parameters and exploring the effects of their variation within ranges consistent with what we know about the CTU agriculture. The nominal values of the input parameters, their dimensions and the mathematical notation used in the equations are given in Table 8, whereas Table 9 contains the most important results of the calculations presented in a similar format. The text contain sufficient detail to reproduce all the results of Table 9, and to calculate any other quantity if it is not given in the latter table.

## 6.1 Per capita cereal production and arable land area

With the daily dietary requirement of $c$ [kcal/person/day], the annual diet must have the calorific value $C = 365c$ [kcal/person/year]. The relative contributions of cereals, domestic animal products and wild animal products to the diet are denoted $\varepsilon_g$, $\varepsilon_d$ and $\varepsilon_w$, respectively (see Section 5). Thus, the annual calorific values of grain (cereals), domestic and wild animal products required for one person to subsist are $\varepsilon_g C$, $\varepsilon_d C$ and $\varepsilon_w C$, respectively.

The cereal yield available for consumption, $Y_g$ [kg/ha/year], is obtained from the total yield $Y$ by subtracting various losses and the amount required for seeding. We assume that a fraction $\gamma$ of the cereal yield is used for seeding and a fraction $\lambda$ of the total grain amount is lost to pests and other losses; the nominal figures are $\gamma = 0.12$ and $\lambda = 0.25$. The usable cereal yield is then $Y_g = (1 - \gamma - \lambda)Y = 0.63Y$. With the calorific value of grain equal to $e_g$ [kcal/kg] and the crop area per person equal to $A_g$ [ha/person], the calorific value of the cereals grown annually is given by

$$E_g = e_g Y_g A_g .$$

The per capita crop area required to satisfy the dietary needs in cereals follows from the requirement that energy produced annually, denoted $E_g$, equals the annual cereal dietary energy requirement, $\varepsilon_g C$:

$$A_g = \frac{\varepsilon_g C}{e_g Y_g} . \qquad (5)$$

However, only a fraction of the arable area is used for the crops, and the rest is fallow; the area of the fallow fields exceeds that under the crops by a factor $\delta_f$. Furthermore, only a fraction $\delta_a$ of the total land area is arable. Thus, the *total* land area containing the cereal fields and fallow land required to satisfy the dietary requirements of a single person is given by

$$A_\text{f} = A_\text{g} \frac{1+\delta_\text{f}}{\delta_\text{a}}. \quad (6)$$

For the sake of simplicity, we assume that there is only one type of cereal (and domestic plants in general) grown for food, but the diversity of crops (including legumes) can easily be allowed for by introducing the dependence of the usable cereal yield $Y_\text{g}$ on the yields and nutrition values of any other cereal varieties and including other cultivated plants into the calorific dietary budget, in the same manner as it is done below for the animal products. We restrain ourselves from including all these factors into our calculations only to avoid any misinterpretation of their accuracy.

## 6.2 Per capita consumption of domestic animal products and the livestock grazing area

A similar calculation for animal food products is slightly more complicated, as there is more than one kind of domestic animals kept and wild animals hunted for. Since the amount of food provided and the grazing area required are rather different for different animals, it is more important to allow explicitly for the herd diversity than for the crop diversity.

The bone assemblages discussed above provide the relative average numbers of cattle, sheep/goat, pig and horse among the domestic animals kept, denoted here $a_\text{c}$, $a_\text{s}$, $a_\text{p}$ and $a_\text{h}$. Their usable meat weight is denoted $m_\text{c}$, $m_\text{s}$, $m_\text{p}$ and $m_\text{h}$, respectively. Consider a herd of this composition that has $n_\text{a}$ animals per capita. Given that fractions $k_\text{c}$ [1/year] of the cattle, $k_\text{s}$ of the sheep/goat and $k_\text{p}$ of pigs are slaughtered for meat, the per capita energy content of the meat procured per year [kcal/year/person] can be calculated as

$$E_\text{a} = n_\text{a} \left( k_\text{c} a_\text{c} m_\text{c} e_\text{c} + k_\text{s} a_\text{s} m_\text{s} e_\text{s} + k_\text{p} a_\text{p} m_\text{p} e_\text{p} + k_\text{h} a_\text{h} m_\text{h} e_\text{h} \right), \quad (7)$$

where individual terms in the brackets represent the contributions of beef, lamb/mutton and pork, respectively. We include horses here for generality, although we will later assume that horses are not kept for food (perhaps as draught animals) and neglect their contribution to the diet. Since the cattle and caprines are kept for both meat and milk, it is reasonable to assume equal cull rates for these animals, $k_\text{c} = k_\text{s}$, but the cull rate of pigs can be larger.

The per capita area $A_\text{a}$ required for the animals to graze is given by

$$A_\text{a} = n_\text{a} \left( a_\text{c} A_\text{c} + a_\text{s} A_\text{s} + a_\text{p} A_\text{pi} + a_\text{h} A_\text{h} \right), \quad (8)$$

where $a_i A_i$ (with $i$ = c, s, pi, h for the cattle, sheep/goats, pigs and horses, respectively) are the proportions of grazing areas of various animals in the total grazing area. The grazing area includes meadows, fallow land and woodland; pigs and goats can find food even near to or within a rural settlement. In calculations presented below, we assumed that pigs do not need any grazing area additional to that used by other animals; formally, we put $A_\text{pi} = 0$.

The per capita area required to collect winter fodder for the livestock is similarly calculated as

$$A_\text{p} = n_\text{a} \left( a_\text{c} M_\text{c} + a_\text{s} M_\text{s} + a_\text{h} M_\text{h} \right),$$

where $M_i$ (with $i$ = c, s, h for the cattle, sheep/goats and horses, respectively) are the land areas required to produce fodder for one head of the corresponding animal.

Table 8. Input parameters related to the palaeodiet and agricultural practice, their dimensions, nominal values and notation. Justification of the nominal values and references can be found in the text.

| Parameter | Unit | Nominal value | Notation |
|---|---|---|---|
| **Dietary requirements** | | | |
| Energy content of the daily diet per person | kcal/person/day | 2500 | $c$ |
| Annual energy content of the diet per person | kcal/person/year | | $C$ |
| Calorific fraction of cereal products in the diet | | 0.4 | $\varepsilon_g$ |
| Calorific fraction of domestic animal products in the diet | | 0.4 | $\varepsilon_d$ |
| Calorific fraction of wild animal products in the diet | | 0.2 | $\varepsilon_w$ |
| **Cereals** | | | |
| Cereal yield | kg/ha/year | 700 | $Y$ |
| Seeding fraction of the yield | | 0.12 | $\gamma$ |
| Crop fraction lost to pests and other losses | | 0.25 | $\lambda$ |
| Energy content of the grain | kcal/kg | 3150 | $e_g$ |
| **Herd composition: relative numbers of various animals, by head** | | | |
| Cattle | | 0.35 | $a_c$ |
| Sheep/goat | | 0.24 | $a_s$ |
| Pig | | 0.33 | $a_p$ |
| Horse | | 0.08 | $a_h$ |
| **Meat of domestic animals and dairy products** | | | |
| Cattle: usable meat weight | kg/head | 100 | $m_c$ |
| Sheep/goat: usable meat weight | kg/head | 25 | $m_s$ |
| Pig: usable meat weight | kg/head | 20 | $m_p$ |
| Horse: usable meat weight | kg/head | 100 | $m_h$ |
| Energy content of beef | kcal/kg | 1600 | $e_c$ |
| Energy content of lamb/mutton | kcal/kg | 1600 | $e_s$ |
| Energy content of pork | kcal/kg | 3000 | $e_p$ |
| Energy content of horse meat | kcal/kg | 1600 | $e_h$ |
| Fraction of milking cows in the cattle herd | | 0.5 | $\kappa_c$ |
| Fraction of milking ewes/does in the caprine herd | | 0.25 | $\kappa_s$ |
| Surplus cow milk yield (after weaning) | litre/year/head | 400 | $y_c$ |
| Surplus sheep/goat milk yield (after weaning) | litre/year/head | 50 | $y_s$ |
| Energy content of cow milk | kcal/litre | 600 | $e_{mc}$ |
| Average energy content of caprine milk | kcal/litre | 800 | $e_{ms}$ |
| **Animal husbandry** | | | |
| Fraction of cattle and sheep/goat killed-off annually | 1/year | 0.2 | $k_c$ |
| Fraction of pigs killed-off annually | 1/year | 0.5 | $k_p$ |
| Grazing area per cow | ha/head | 10 | $A_c$ |
| Grazing area per sheep/goat | ha/head | 0.5 | $A_s$ |
| Area for winter cattle/horse fodder per head | ha/head | 0.3 | $M_c$ |
| Area for winter sheep/goat fodder per head | ha/head | 0.02 | $M_s$ |
| Fraction of leafy fodder | | 0.5 | |
| **Wild animal products** | | | |
| Red deer usable meat weight | kg/head | 130 | $m_r$ |
| Roe deer usable meat weight | kg/head | 11 | $m_{ro}$ |
| Wild boar usable meat weight | kg/head | 130 | $m_b$ |
| Fraction of red deer among hunted animals, by head | | 0.47 | |
| Fraction of roe deer among hunted animals, by head | | 0.24 | |
| Fraction of wild boar among hunted animals, by head | | 0.29 | |
| Energy content of red deer meat | kcal/kg | 1400 | $e_r$ |
| Energy content of roe deer meat | kcal/kg | 1400 | $e_{ro}$ |
| Energy content of wild boar meat | kcal/kg | 3500 | $e_b$ |
| **Labour productivity** | | | |
| Fraction of a farming family fit to work in the fields | | 0.33 | $w$ |
| Area tilled with hand tools by one person in one hour | m²/person-hour | 15 | $s_1$ |
| Area ploughed with ard by one person in one hour | m²/person-hour | 260 | $s_1$ |
| Crops or grass area reaped by one person in one hour | m²/person-hour | 30 | $s_2$ |
| Crops area threshed and winnowed by one person in one hour | m²/person-hour | 30 | $s_3$ |
| Length of a working day | hours/day | 10 | $\tau$ |

| Land use and the settlement exploitation area | | | |
|---|---|---|---|
| Arable fraction of the land area | | 0.35 | $\delta_a$ |
| Ratio of the fallow field to the crops area | | 2 | $\delta_f$ |
| Ratio of the agriculturally-unproductive to the total land area | | 0.2 | $\delta_u$ |
| Fraction of the land area producing winter fodder | | 0.5 | |
| Settlement 1: area | ha | 2 | $A_0$ |
| Settlement 2: area | ha | 10 | $A_0$ |
| Population density within a settlement | person/ha | 27 | |

Table 9. Results of calculations related to the palaeodiet and agricultural practice for the nominal parameter values given in Table 8.

| Variable | Unit | Nominal value | Notation |
|---|---|---|---|
| **Cereals** | | | |
| Wheat yield available for consumption | kg/ha/year | 441 | $Y_g$ |
| Per capita daily amount of cereals consumed | kg/person/day | 0.32 | |
| **Per capita herd composition and land use** | | | |
| Herd size: the total number of animals | head/person | 5.2 | $n_a$ |
| Cattle | head/person | 1.8 | $n_c$ |
| Sheep/goat | head/person | 1.3 | |
| Pig | head/person | 1.7 | |
| Horse | head/person | 0.4 | |
| Grazing area | ha/person | 22.8 | |
| Fodder area | ha/person | 0.7 | |
| **Per capita daily consumption of domestic meat and dairy products** | | | |
| Beef | kg/person/day | 0.1 | |
| Cow milk | litres/person/day | 1.0 | |
| Lamb/kid/mutton | kg/person/day | 0.02 | |
| Caprine milk | litres/person/day | 0.04 | |
| Pork | kg/person/day | 0.05 | |
| **Per capita daily consumption of wild animal products** | | | |
| Total number of wild animals hunted annually | head/person/year | 0.8 | |
| Read deer hunted annually | head/person/year | 0.4 | |
| Roe deer hunted annually | head/person/year | 0.2 | |
| Wild boar hunted annually | head/person/year | 0.2 | |
| Red deer meat, daily consumption | kg/person/day | 0.138 | |
| Roe deer meat, daily consumption | kg/person/day | 0.006 | |
| Wild boar meat, daily consumption | kg/person/day | 0.085 | |
| **Per capita labour costs and labour return** | | | |
| Tillage with hand tools | person-day/person | 18 | |
| Tillage with ard | person-day/person | 1 | |
| Reaping, threshing and winnowing | person-day/person | 18 | |
| Grass cutting for winter fodder | person-day/person | 12 | |
| Labour return: ratio of the total time available to working time: | | | $\eta$ |
|     tilling with hand tools | | 7.8 | |
|     ploughing | | 12.2 | |
| **Per capita land use and the settlement exploitation area** | | | |
| Crops area | ha/person | 0.26 | $A_g$ |
| Fallow area | ha/person | 0.53 | |
| Grazing area (in addition to fallow land) | ha/person | 1.01 | $A_p$ |
| Field zone area | ha/person | 2.25 | |
| Grazing zone area | ha/person | 26.63 | |
| Fodder zone area | ha/person | 0.69 | |
| Local subsistence carrying capacity | persons/km$^2$ | 3.4 | $K_s$ |
| Settlement 1: | | | |
|     Settlement radius | km | 0.08 | $R_0$ |
|     Maximum distance to the field zone | km | 0.54 | $D_1$ |
|     Maximum distance to the grazing zone | km | 2.16 | $D_2$ |
|     Maximum distance to the fodder zone | km | 2.21 | $D_3$ |
| Settlement 2: | | | |
| Settlement radius | km | 0.18 | $R_0$ |
| Maximum distance to field zone | km | 1.22 | $D_1$ |
| Maximum distance to the grazing zone | km | 4.84 | $D_2$ |
| Maximum distance to the fodder zone | km | 4.95 | $D_3$ |

A perhaps unexpected result of our calculations (confirmed by Jorgenson 2009) is that dairy products can play quite a significant and important role in the diet. With the calorific values of the cow and caprine milk denoted $e_{mc}$ and $e_{ms}$, and the respective per capita animal numbers given by $n_a a_c$ and $n_a a_s$, the per capita amount of milk that can be obtained annually from the herd [litre/year/person] is given by

$$Y_m = n_a \left( \kappa_c a_c y_c + \kappa_s a_s y_s \right),$$

where $\kappa_c$ and $\kappa_s$ are the fractions of milk-producing cows and caprines in the herd. Having in mind the limited accuracy of any estimates of this kind, we neglect the relatively small number of the male cattle in the herd and thus assume that the value of $a_c$ is the same here and in Eq. (7) for the meat production and Eq. (8) for the grazing area. We allow for the fact that only a fraction of the cows, ewes and does can be milked at any time. The lactation period of cow is close enough to half a year ranging from 180 to 230 days (Gregg, 1988, p. 106); thus, we adopt $\kappa_c = 0.5$. The lactation period of unimproved breed of caprines varies from 12 weeks (Banda et al. 1992) to 19 weeks for sheep and 30 weeks for goats (Redding 1981, cited in Gregg, 1988, p. 116). We adopt the lower value, 12 weeks annually, to have $\kappa_s = 0.25$ but the range 0.25–0.5 appears to be a realistic possibility.

Analyses of archaeological bone assemblages do not always distinguish between the sheep and the goat. This may affect the estimate of the energy content of the dairy products since the energy content of the cow milk, about $e_{mc} = 600$ kcal/litre on average, differs significantly from that of the sheep milk, 1030 kcal/litre, but not the goat milk, 680 kcal/litre (Table 3.1 of Muchlhoff et al. 2013). We adopt the energy content of the caprine milk at about the average of the latter two figures at $e_{ms} = 800$ kcal/litre. Then the energy content of the milk available annually from $n_a$ animals follows as

$$E_m = n_a \left( \kappa_c a_c y_c e_{mc} + \kappa_s a_s y_s e_{ms} \right). \tag{9}$$

We assume that all this energy is consumed in the form of various dairy products if not milk itself.

Equating the total calorific value of the meat and dairy products obtained from the herd, $E_a + E_m$ from Eqs. (7) and (9), to the calorific value of domesticated animal products required to satisfy the dietary requirements of one person, $\varepsilon_d C$ [kcal/person/year], we obtain $n_a$, the number of animals in the herd required to satisfy the dietary requirements of a single person in animal products:

$$n_a = \frac{\varepsilon_d C}{a_c k_c m_c e_c} \left[ 1 + \frac{k_s a_s m_s e_s}{k_c a_c m_c e_c} + \frac{k_p a_p m_p e_p}{k_c a_c m_c e_c} + \frac{k_h a_h m_h e_h}{k_c a_c m_c e_c} + \frac{\kappa_c y_c e_{mc}}{k_c m_c e_c} \left( 1 + \frac{\kappa_s a_s y_s e_{ms}}{\kappa_c a_c y_c e_{mc}} \right) \right]^{-1}. \tag{10}$$

Assuming that horses are not used for food (in part, because of their relatively small numbers relative to the cattle), we neglect their contribution to meat supply, formally setting $m_h = 0$ in the above equations. This is consistent with the fact that the relative number of horses increases in the Late Trypillia (Table 7), as the need in draught animals is likely to increase as agriculture becomes more intensive. The grazing area required for the pigs is neglected, $A_{pi} = 0$ (see Section 5.2).

The number of domestic animals required to satisfy dietary requirements of $N$ people can be calculated as

$$N_a = N n_a .$$

Then the numbers of the cattle, caprines, pigs and horses in the herd are equal to $N_a a_c$, $N_a a_s$, $N_a a_p$ and $N_a a_h$, respectively.

## 6.3 Wild animal products

The final contribution to the calorific value of the palaeodiet considered here comes from the meat of wild animals, red deer, roe deer and wild boar. As discussed in Section 5.2, bone assemblages at the CTU as well as other Neolithic and Bronze Age sites suggest that about $\varepsilon_w = 0.2$ of the total energy intake was from the wild animal meat. Using their relative numbers, meat weight and calorific values given in Table 7, one can convert the required energy content into the numbers of the wild animals per person implied by the bone assemblages in the same way as is done for domesticated animals. We do not write out these relations here since they differ insignificantly from those already given.

## 6.4 Per capita subsistence land area and the subsistence carrying capacity

The total land area required to provide the amounts of cereals, meat and dairy products of domestic animals to satisfy the calorific dietary requirements of a single person is the sum of the specific land areas under cereals and pasture obtained in Sections 6.1 and 0:

$$A = A_a + A_p ,$$

and the local subsistence carrying capacity $K_s$ [persons/km$^2$] follows as

$$K_s = \frac{1}{A} .$$

This estimate needs careful qualification to be useful. Although $K_s$ is called here a carrying capacity, it should not be confused with the maximum population density averaged over a large area that appears in demographic and population dynamics models. It is based on the land area required to support a single person and is used below to calculate the area required to support a rural settlement (the exploitation territory). However, the exploitation areas of settlements do not need and, indeed, are unlikely to cover the landscape completely while the land between the exploitation areas does not enter our calculations. Therefore, $K_s$ represents the upper limit of the carrying capacity, attainable only under an unrealistic condition of densely packed exploitation areas. To extend such calculation to the global carrying capacity, careful analysis of the spatial patterns and lifetimes of the settlements is required as well as detailed environmental data. An example of such analysis can be found in Zimmermann et al. (2009) who suggest 8.5 persons/km$^2$ for the local carrying capacity of LBK settlements in 5250–5050 BC and note its strong spatial variability, whereas their global estimate is 0.6 persons/km$^2$. Ellen (1982, p. 43) notes that the actual population densities are most often well below the local carrying capacity at a level of 25–70%.

## 6.5 The maximum fraction of manured fields

The above relationships between rainfall, duration of cultivation and yield can be used to estimate the average yield at the CTU sites with allowance for the use of manure fertiliser. The overall yield $Y$ [kg/ha/year], given the fraction of manured land, $f_m$, is given by

$$Y = Y_m f_m + Y_u (1 - f_m), \tag{11}$$

where $Y_u$ and $Y_m$ are the yields from unmanured and manured fields, respectively. The amount of manure available depends on the amount of livestock kept and its management. The finds of faunal remains at the CTU sites (Table 7) can be used to estimate the *maximum amount* of manure which could be used as a fertiliser. To estimate $f_m$, we use the following variables: $\mu$, the amount of manure applied [kg/ha/year]; $m$, the amount of manure collected per head of cattle [kg/head/year]; $n_c$, the number of cattle kept per person [head/person]; and $\varepsilon_g C$, the consumption of wheat per person [kg/person/year]. We will not be counting manure in the same detail as the meat, milk and grazing area, although it is easy to do, and will only include cattle manure into the calculation. Then the maximum fraction of the manured arable land, attainable if *all* the manure produced is used in the fields, is estimated as

$$f_m = \frac{m n_c}{\mu (1 + \delta_f) A_w},$$

where $A_w$ is the crops area per person from Eq. (5) and $(1 + \delta_f) A_w$ is the total area of both cropped and fallow fields. Using Eq. (11) for $Y$ in Eq. (5), with $Y_w = (1 - \gamma - \lambda)$, we obtain a simple equation for $f_m$, which solves to yield

$$f_m = \frac{1}{1 - \dfrac{Y_m}{Y_u} + \dfrac{C \mu \varepsilon_g (1 + \delta_f)}{m n_c e_g Y_u (1 - \gamma - \lambda)}}. \tag{12}$$

We take $\mu = 15$ tonne/ha/year, as in the Sanborn experiments, and $m = 2.5$ tonne/head/year for the manure from cattle (LWFH 1993), assuming that 50% of the total amount of the manure is lost. The Sanborn data on wheat yields from manured and unmanured plots, summarized in Eq. (4) and Table 6, suggest $Y_m/Y_u = 1.2$ for $D = 10$ years and the typical rainfall in the CTU area, $R = 550$ mm/year. Assuming $Y_u = 650$ kg/ha/year, with the nominal per capita cattle number $n_c = 1.8$ head/person and other variables from Table 9, we obtain $f_m \approx 0.4$, that is, about half of the total field area (both cropped and fallow) could be fertilised with the manure available for the nominal values of the parameters. Using Eq. (11), we then obtain the nominal average yield of $Y_0 = 700$ kg/ha/year (in fact, we have adjusted the above value of $Y_u$ to preserve consistency with the nominal parameter values of TablesTable 8 andTable 9).

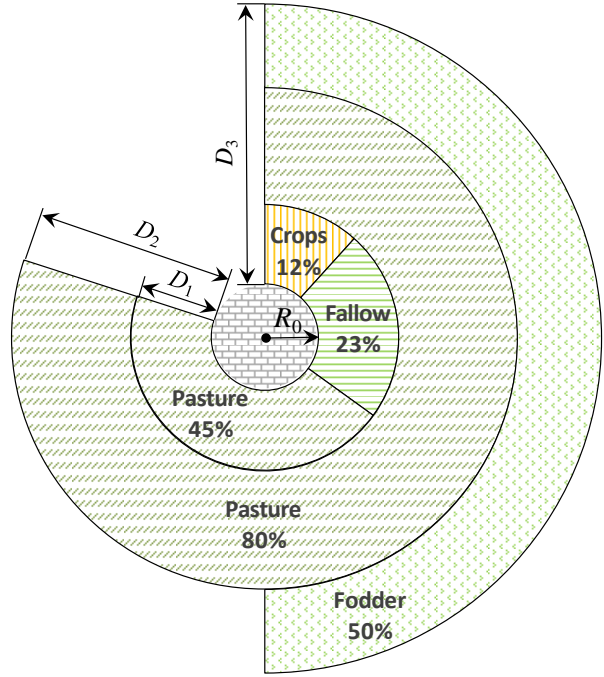

**Figure 2.** Land use of a settlement for the nominal diet structure with the relative fractions of cereals, domestic and wild animal products of 0.4, 0.4 and 0.2, respectively, and the cereal yield of 700 kg/ha/year. The settlement is represented by the innermost circle, surrounded by the field zone containing the area under crops (12%), fallow fields (23%) used for pasture, and specialized grazing area (45%), leaving 20% of the area for unproductive land (unshaded – ravines, dense forests, etc.). The next outer zone is used exclusively for livestock grazing; it also contains 20% of the area that cannot be used for any agricultural purposes. The outermost zone is used to collect animals' winter fodder from both grass meadows and suitable trees that are assumed to occupy a half of the total area in that zone. The settlement radius $R_0$ and the maximum distances to the zones, $D_1$, $D_2$ and $D_3$ (shown not to scale), are discussed in the text and given in Table 9.

## 7. The exploitation territory of a settlement

With the above estimates, we can calculate the land area exploited by the population of a rural settlement. Consider a settlement of an area $A_0$ with a population of $N$ people. Here and below, we assume for simplicity that the settlement area is circular, so that $A_0 = \pi R_0^2$, where $R_0$ is its radius. In fact, many CTU settlements have a roughly elliptical shape; then $R_0$ is understood as the geometric mean of the minor and major semi-axes of the settlement, $r_1$ and $r_2$: $R_0^2 = r_1 r_2$.

The land around a settlement is thus divided into three zones shown in Figure 2. The field zone is the closest to the settlement, where both currently cultivated and fallow fields are located. Fallow fields in this zone can be used for grazing. The next outer zone is used as summer pasture for the livestock. The outermost zone is where winter fodder for the animals is collected. The total area of the field zone serving $N$ people is given by

$$A_1 = NA_f,$$

with the per capita area $A_f$ given in Eq. (6), and contains the crops area $NA_g$ and fallow land of an area $N\delta_f A_g$; the remaining land is agriculturally unproductive. Most of the pasture and grazing areas are located in the grazing zone at a larger distance from the settlement. The fallow area $N\delta_f A_g$ in the field zone can be used for grazing, so that the useful area of the grazing zone has to be equal to $NA_a - N\delta_f A_g$, where $NA_a$ is the total grazing area required, with $A_a$ given in Eq. (8).

The total area of the grazing zone (including unproductive land of the same fractional area $\delta_u$) is then given by

$$A_2 = N \frac{A_a - \delta_f A_g}{1 - \delta_u}.$$

Finally, the area of the fodder zone serving $N_a$ animals is given by $A_m = N_a(a_c M_c + a_s M_s + a_h M_c)$, and its total area follows as

$$A_3 = \frac{A_m}{\delta_m},$$

where $\delta_m$ is the fraction of the total area bearing meadows and trees providing leafy fodder; we adopt, more or less arbitrarily, $\delta_m = 0.5$. Since the radius of the fodder zone is relatively large (Section 7), the magnitude of $\delta_m$ affects the radius of the zone only slightly: a change of $\delta_m$ form 0.1 to 0.9 changes by just 10% the outer radius of the fodder zone around a settlement with 2000 inhabitants.

It is straightforward to calculate the radial distances to the boundaries of the three exploitation zones from either the centre of a settlement or its border assuming that they have circular shape. For example the maximum distance from the settlement border to the fodder zone is given by

$$D_3 = \sqrt{(A_0 + A_1 + A_2 + A_3)/\pi} - R_0,$$

and similarly for the maximum distance to the field and pasture zones, $D_1$ and $D_2$, respectively.

## 8. Labour costs of the agricultural cycle

For the estimates described above to be viable, one has to demonstrate that the food required can indeed be produced with the labour resources available. The availability of human labour rather than land could be the limiting factor in the early agriculture (Halstead 1996); our calculations confirm this. In this section, we discuss the labour required for a farming population to subsist, starting with estimates of labour productivity in pre-modern agriculture and proceeding to evaluating the labour costs of the agricultural cycle and then, the labour efficiency. Knowing the area required for the population, we then estimate its local subsistence carrying capacity within the exploitation area.

### 8.1 Experiments on agricultural labour productivity

Archaeological finds at CTU sites include a range of agricultural tools, including stone and antler hoes and flint sickle blades; remarkably, an antler ard was found at Grebenukiv Yar (near Maydanetske), dated to the late-fifth–early-fourth millennium BC (Pashkevich and Videiko 2006, pp. 88–95). Numerous ceramic models of sledges with ox heads clearly suggest the use of cattle for traction (Pashkevich and Videiko 2006, p. 89), confirming the conclusions drawn from analyses of faunal remains (Zhuravlev 2008).

Semyonov (1974, pp. 194–226) describes in detail extensive experiments conducted in 1969–1970 at the Laboratory of Primitive Techniques in the Leningrad Branch of the Institute of Archaeology of the Academy of Science of the USSR. The experiments involved tilling and harvesting with tools modelled upon prehistoric and ethnographic examples. The tools tested include various digging sticks, stone, wood and antler hoes, wooden ards, and sickles with flint blades. In those experiments, friable soil could be prepared for sowing (tilled to a depth of 20–25 cm) with an oak dibble at a rate ranging from $s_t = 50$ m²/person-hour on a well-manured field to 10 m²/person-hour on a denser soil and to 5 m²/person-hour on a dense, half-virgin soil. Adding an iron point to an oak dibble increased the productivity to $s_t$ = 6–8 m²/person-hour on virgin soil, and to $s_t$ = 8–15 m²/person-hour when the stick was further equipped with a pedal. Work with a dibble with an additional weight was slightly more productive but required a significantly larger physical effort. With a stone hoe, $s_t$ = 13–17 m²/person-hour of light soil could be tilled, somewhat better than with an antler hoe at $s_t$ = 6–17 m²/person-hour. Tilling of virgin soil covered with high grass and dense turf could be done at a rate $s_t$ = 2.5 m²/person-hour with an oak dibble and about $s_t$ = 6 m²/person-hour with hoes (2 hours 10 min of work with an antler hoe followed by 1 hour 15 min using an iron hoe on a plot 25 m² in size). Altogether, the productivity of hand tilling with a digging stick or stone hoe can be adopted as $s_t$ = 10–20 m²/hour depending on the soil quality.

Tilling with horse-drawn oak ards, modelled on the earliest prehistoric evidence, involved two people, one to guide the horse and the other to manipulate the ard. A plot 250 m² in size, with soil tilled earlier but hardened after a 12-day drought, could be tilled with a Døstrup (spade) ard in 40 minutes (375 m²/hour) to the depth of 30–35 cm, whereas tilling a similar plot on the same field with digging sticks and hoes took about 50 hours. Thus, the tillage efficiency is increased by more than a factor of 50. Cross-ploughing of the plot with a Walle (crook) ard was equally successful. However, both ards failed to perform on virgin soil covered with grass. The Walle ard was tested on a previously harvested pea–oat field with stubble, plant roots and weed on dry soil compressed by the heavy machinery used for harvesting. An area of 1430 m² was tilled to a depth of 10–20 cm in 2 hours 50 minutes (about 500 m²/hour). Although the depth of tilling with hand tools was 1.5–2 times larger and the furrows made with the ard were unevenly spaced, the soil tilled with the ard was better pulverised. Cross-ploughing of the plot removed the imperfections in additional 2 hours 35 min. Trials of the Døstrup ard on a clayey soil after a strong rain demonstrated the difficulties of working on sticky soil with higher resistance from wet plant roots and weeds. A single ploughing of 1430 m² took 3 hours 20 minutes (about 430 m²/hour) in this case. Altogether, ploughing 1430 m² twice by two people took 5 hours 25 minutes, or at the overall rate of about $s_t$ = 260 m²/person-hour. We note in passing that of the two workers involved in ploughing, a physically weaker person, e.g., an older child, can guide the animal.

Semyonov (1974, p. 252) cites Steensberg (1943, pp. 10–22) who experimented with harvesting ripe barley and partially ripe oats with modern and primitive sickles in 1938–1939 near Lviv (Lemberg) in Western Ukraine, in Slovakia and in Denmark. With a flint sickle, cutting low on the stem at a height of 12–30 cm above ground was done at a rate $s_r$ = 30–40 m²/person-hour (10 m²/person-hour are equivalent to 100 person-day/ha for a 10-hour working day). Mowing of 50 m² with a Viking- or Roman-type scythe took 17–30 minutes. Semyonov's (1974, pp. 253–254) own experiments on cutting wet grass (stem diameter 0.5–0.7 mm) with flint sickles, modelled on those found at the CTU site Luka-Vrublevets'ka, resulted in a $s_r$ = 20–25 m²/person-hour productivity; ripe rye could be reaped at a slightly higher rate, $s_r$ = 20–35 m²/person-hour. A cultivated fodder field (oats, barley, peas, goose-foot and 10% of various weeds, up to 1.5 m in height and 0.8 cm in the stem diameter) could be reaped (by cutting the stem at a height of 25 cm above ground or larger) with flint sickles at a rate 20–30 m²/person-hour. Altogether, Semyonov (1974, pp. 255–256) concludes that the productivity of reaping with a flint sickle is only twice lower than with a modern steel tool.

White (1965) assesses as credible Columella's estimates of the average labour cost for Roman Italy to be about 44 person-day/ha (18 person-day/acre) for the whole wheat cultivation cycle, excluding harvesting, with four ploughings (including ploughing-in the seed), and further 5.7 person-day/ha (1.5 person-day/iugerum) for reaping. Halstead and Jones (1989) describe traditional farming in modern Greek islands. Their conclusions emphasize the highly seasonal nature of agricultural activity with maximum time stress in the harvesting period and, to a lesser extent, the ploughing season. These authors also note that overproduction and storage of more than one year's supply of food is a relevant response to the risk of a failing crop inherent in a highly seasonal climate environment. A typical labour cost of reaping cereals with a modern sickle was 10–30 person-days/ha, and the crop processing (threshing, winnowing, etc.) required about the same amount of labour as the reaping. Assuming 0.75–1.2 ha/person of per capita cultivated land, harvesting at this rate would take 7.5–36 person-days/person. A typical modern productivity of tilling is 25 m$^2$/hour (1 ha in 400 hours) when using hand tools, and about 150 m$^2$/hour (1 ha in 65 hours) when tilling with a pair of oxen (Ellen 1982, p. 137).

## 8.2 The agricultural cycle and labour return

Using estimates of the labour productivity presented above, the dietary requirements presented in Sections 3 and the land use estimates of Section 6, it is straightforward to estimate the labour cost of the arable farming and livestock maintenance required for the population to subsist.

Equation (5) expresses the area under crops in terms of the per capita dietary requirements in cereals and the cereal yield. Using the nominal values of the labour productivity presented in Table 9, we obtain the estimates of the labour cost of various agricultural activities collected under the *Labour Productivity* heading in that table. Whenever required, we assumed that a working year consists of 250 days, allowing for bad weather, holidays, etc. (White 1965).

It is convenient to summarise some (but not all) important aspects of the organization of farming in terms of the labour return, denoted $\eta$, which can be defined as the ratio of energy produced to the energy spent or, equivalently, as the ratio of the length of time over which a person can subsist (here, in terms of the calorific food content alone) on the food produced, to the working time required to produce it. Based on ethnographic evidence, Ellen (1982, p. 45) suggests that an overall labour return of 10 is about the minimum acceptable in subsistence societies, with 1750 kcal produced per person-hour of labour for major economic activities. However, the labour return of plant cultivation alone can be as low as $\eta = 2.4$ among swidden horticulturalists in modern Indonesia (Ellen 1982, p. 152) To illustrate the significance of this quantity, we note that, theoretically, one person can support themselves with the labour return of at least unity; to support a family of six, two working family members must achieve a labour return of at least three; if any surplus food should be produced, as an emergency storage or for exchange, higher labour return is required.

In our calculations, we focus on the costs of labour that requires a certain physical fitness, such as land tillage, and on those seasonal activities that must be completed in a limited time, such as land preparation for sowing and reaping the harvest and winter fodder. These are the most demanding parts of the agricultural cycle in terms of either the workforce or time. We assume that only a fraction $w$ of a family members are capable of physically demanding work, with $w = 1/3–1/2$. Many other activities, such as sowing, cleaning the grain, collecting leafy fodder, can be assigned to less capable family members and/or spread over longer time.

Even at the lower-end tillage productivity of 15 m$^2$/person-hour, it takes only 66 person-days to satisfy the annual dietary requirements of a single person in cereals. Considering a family of six people of whom only two are physically fit to work ($w = 1/3$), the cost of producing the grain required for its annual subsistence is just 396 person-days per family per year, as compared to 500 man/days available annually in such a family. The resulting labour return is reasonably high at $\eta = 365$ person-days/66 person-days $\approx 5.5$.

However, a problem with this option is that the tilling of a family cereal field requires 104 person-days, or 52 days if done by two workers, while the soil preparation and sowing must be done in not more than 30 days to avoid significant crop losses (Percival 1974, p. 423). Tilling the family field with hand tools by two people can only be finished in 31 days if the productivity is $s_t \approx 25$ m$^2$/person-hour. This is marginally acceptable but still leaves little room for any eventualities such as bad weather or difficult soil. There are several ways to resolve the problem. An obvious one is to have more family members working in the fields, especially during the tillage and sowing. For half of the family members tilling the field at a rate $s_t \approx 15$ m$^2$/person-hour, the work can be finished in about 31 days. Another obvious option could be to use wheat varieties that provide higher yield. However, this does not lead to any significant saving in the labour. For example, two people working at $s_t \approx 15$ m$^2$/person-hour could till the family field in 30 days only provided the wheat yield was as implausibly high as $Y = 2200$ kg/ha/year (with a high labour return of about nine, though). Neither winter crops nor manuring alone is likely to boost the yield to that level. Yet another option is to reduce the reliance on cereals by reducing their contribution to the diet. This could be achieved, for instance, if 20% of the calorific content of the diet was from cereals and 60% from domestic animal products, provided the cereal yield is $Y = 1100$ kg/ha/year.

A more radical, and long-term, solution is to replace the hoe with the ard. Then two workers can plough the family field in just 3 days. As mentioned above, primitive ards are not efficient on heavy and virgin soils where the hoe appears to be the only alternative. This fact highlights the difficulty of moving the fields to a virgin soil if the settlement has to be relocated.

Another bottleneck in the agricultural cycle is cutting grass for the winter fodder. If only meadow grass was used for fodder, working with a flint sickle would require 114 person-days to provide the family livestock for winter. This is obviously untenable, even for three workers. However, leaves of certain tree species also provide excellent fodder (see above), and younger or weaker members of a farmer's family could collect them. We assume, admittedly arbitrarily, that only a half of the fodder required is meadow hay. Then the labour cost of fodder (excluding collecting the leaves) is quite acceptable at 67 person-days. A further labour saving option is to improve technology and cut grass with scythe.

There are innumerable such combinations reflecting various techniques and strategies of farming, and there is no point in trying to discuss them all. The diversity in the implementation of farming strategy between individual CTU sites and between CTU evolutionary stages apparent from archaeological evidence is likely to reflect the wide breadth of possibilities. Instead of discussing a large number of hypothetical scenarios, we present our results in a graphical form to show the dependence of the labour return on the wheat yield, the diet structure, etc., with the aim of identifying the limiting elements of a farming strategy. To make the results mutually comparable, we only vary one or a few parameters at a time, having the others fixed at their nominal values given in Table 8.

## 8.3 Trends in the labour return and land use

Calculations of the labour costs of various agricultural activities readily identify the well-known seasonal labour bottlenecks in the farmer's year (e.g., Fuller et al. 2010) where large parts of the annual work have to be done in a limited time: preparation of the land for sowing, collection of winter fodder, and harvesting. The land tilling time, limited to about 30 days, can be an especially demanding constraint. Depending on weather, harvesting may need to be completed in a few weeks or even a few days when the spikelets have not yet dried and shattered. However, this limits mostly the reaping time since the grain can be threshed and cleaned later. Since naked wheat grains are easily detached from the ear, they are better threshed immediately after reaping. On the other hand, hulled wheats can be reaped and then stored to be threshed on daily basis. Thus, we focus on the reaping time in our assessment of the labour costs. Collecting hay, straw and leaves for winter fodder is another activity that may impose stringent time limits. However, younger and weaker members of the family can be involved, relieving pressure on those fit for hard physical work (this is also true of crops reaping). Land tilling thus appears to be the most demanding seasonal activity in terms of the time and labour stress.

To illustrate the results of the calculations, we present per capita figures, e.g., the labour cost of producing enough food to support a single person. Furthermore, we discuss the requirements, and how they could be met, of a family of six people of whom only two or three ($w = 1/3$ or $1/2$) are capable of doing work that require a certain physical fitness. To support such a family, the labour return of two workers must be equal to at least three if $w = 1/3$, or three workers should work with a return of at least two if $w = 1/2$. Finally, we discuss the limiting factors in the agricultural cycle of typical settlements of 2 and 10 ha in area that host about 50 and 270 inhabitants, respectively (assuming a constant population density within the settlements).

Conclusions drawn from the calculations presented further in this section are testable with relevant archaeological material and its analysis. In general, our results imply that certain types of the temporal evolution of the diet are more advantageous and efficient than others, and that different stages in the development of agriculture can have different preferable subsistence strategies.

### 8.3.1 Cereal yield and agricultural technology

One of the constraints on the size of the exploitation area of a rural village is that its fields should be within 5 km at most. This constraint can safely be satisfied even for a large settlement of 75 ha in area as long as the cereal yield exceeds about 350 kg/ha/year.

Figure 3 illustrates a strong effect of the cereal yield on the labour return, for land tillage with either hand tools or ard. Unsurprisingly, the use of ard reduces the labour costs and increases the labour return rather dramatically, by a factor 1.5–2 over the whole agricultural cycle. For a given diet structure, lower yields require larger field areas and, consequently, a larger distance to them. For a settlement of $N = 53$ people and $A_0 = 2$ ha in size, the maximum distance to the crops from the settlement border varies from $D_1 = 0.7$ km to 0.4 km as $Y$ increases from 500 to 1500 kg/ha/year. The maximum distance to the grazing zone varies very little remaining about $D_2 = 2.2$ km; the maximum distance to the fodder zone, $D_3$, differs from $D_2$ by just 50 m. The labour cost of the cereal production varies with the size of the cultivated fields. With 40% of the diet's calorific content coming from cereals, yields below about 400 kg/ha/year are untenable as the amount of labour required to till the land required to feed one person exceeds 31 person-days using hand tools. For a family of six, yields in excess of 1230 kg/ha/year are required to till the family plot in less than 60 person-days; this is just acceptable if two members of the family are fit for hard physical work. Thus, land tillage causes time stress if done with hand tools.

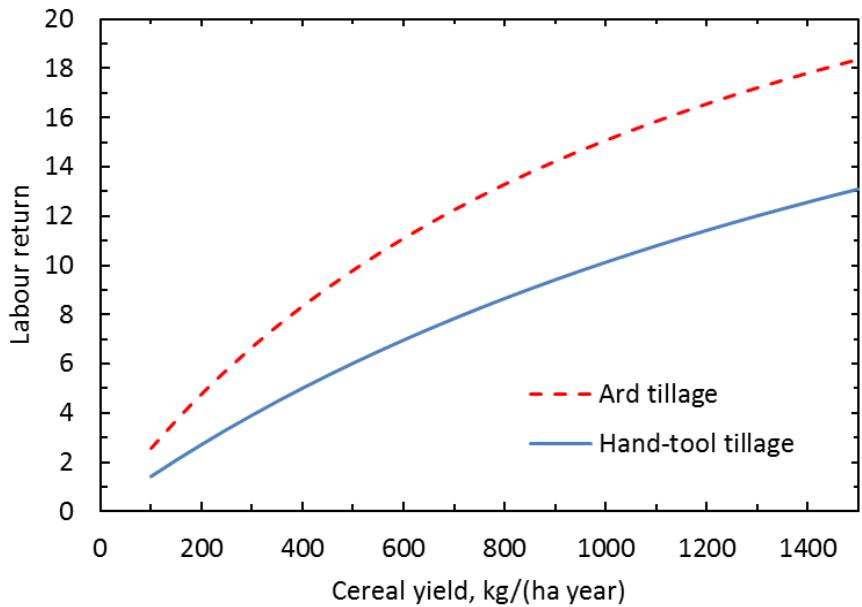

Figure 3. The effect of the cereal yield and the tillage technology on the annual labour return $\eta$, i.e., the amount of output as a fraction of nutritional energy requirement per working hour (or the ratio of the energy produced to the energy spent on the production, or the inverse ratio of the working time required to produce food to the time the output can sustain the worker). Solid (blue): tillage with hand tools; dashed (red): tillage with ard. The diet structure is assumed to be fixed at $\varepsilon_g/\varepsilon_d/\varepsilon_w = 0.4/0.4/0.2$ for the relative contributions of the cereals, domestic and wild animal products.

The use of the ard removes this constraint and leaves abundant time to continue using hand tools, say, in vegetable gardens. Even for implausibly low yields of $Y < 150$ kg/ha/year, the labour required to till a family plot is just 28 person-days.

However, the earliest ard found in the CTU area dates to Trypillia BI. The earliest CTU framers most probably used only hand tools. One option to avoid the excessive time stress in the land tilling and sowing season is to reduce the contribution of cereals to the diet. If the relative contributions of cereals, domestic and wild animal products were $\varepsilon_g/\varepsilon_d/\varepsilon_w = 0.23/0.57/0.2$ (instead of the nominal 0.4/0.4/0.2), the per capita crop area reduces to 0.15 ha/person for $Y = 700$ kg/ha/year, and its tilling would take 10 person-day/person. The labour to prepare a family plot for sowing is, correspondingly, 60 person-day/family. Keeping the livestock is more efficient in terms of the energy return: with the diet containing only 23% of cereals, the labour return is as high as 10. Cutting grass for the herd requires 96 person-day/family; this is a large load but not untenable given that fodder can be collected between the sowing and harvesting seasons by virtually all family members. A possible problem with this option is not in the labour cost but in the distance to the grazing area as the large herd needs a large area to be fed. The distance to the outer boundary of the grazing area around a settlement of about 50 inhabitants (2 ha in area) is $D_2 = 2.6$ km. Larger settlements become still more problematic. For instance, a 10-ha village of about 270 people has its fields within $D_1 = 0.9$ km from the settlement but the outer border of the grazing area is $D_2 = 5.8$ km away. The distance to the fodder zone differs insignificantly (by about 200 m) from that to the grazing area.

The magnitude of $D_2$ obviously depends on the grazing area per animal head while our nominal figure of $A_c = 10$ ha/head is rather generous. Given that less than 1 ha/head of a flooded pasture is sufficient for cattle, $D_2$ can be reduced to 3.8 km for a village of 10 ha in area if $A_c = 5$ ha/head, corresponding to an approximately equal split between meadow and forest grazing (and all other parameters unchanged). With $A_c = 5$ ha/head, a settlement of 20 ha in area still has $D_2 \approx 5.8$ km, but the problem arises again for larger settlements.

Since arable fields represent a relatively small fraction of the exploitation area, changes in the cereal yield affect the local carrying capacity only weakly. As $Y$ varies from 500 to 1500 kg/ha/year for the nominal diet structure, $K_s$ varies by a few percent remaining close to 3.4 persons/km$^2$. Changing the diet to $\varepsilon_g/\varepsilon_d/\varepsilon_w = 0.23/0.57/0.20$ leads to $K_s \approx 2.4$ persons/km$^2$.

Altogether, we suggest that large, exclusively farming settlements of a few thousand people and a few hundred hectares in area are sustainable only if the ard is available. Otherwise, such a settlement has to be supported by satellite farming villages, which would imply complex social organization, labour and occupation division, and well-established, stable exchange networks. The development of complex structures based on technological advances is implausible at the early stages of the CTU. This can be a reason for the dominance of smaller and medium-size settlements in the early CTU. Large, proto-urban settlements have to be supported by adequate technology and/or the developed social relations that presumably emerged at the later stages.

### 8.3.2 The diet structure and labour return

Having identified and quantified specific mechanisms of the influence of the population diet on agricultural activities, we explore this connection in more detail. It appears that reducing the fraction of cereals in the diet is the only obvious way to cope with the labour bottlenecks in a crop-based agriculture, especially if the cereal yield is low.

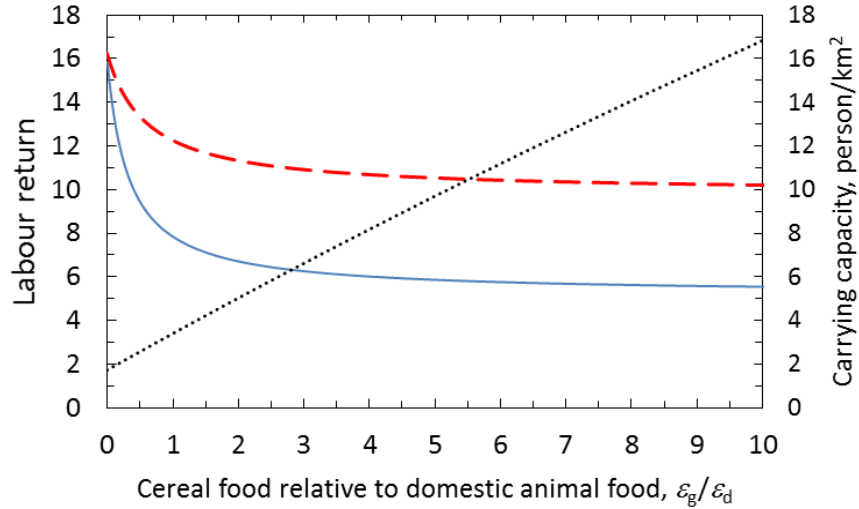

Figure 4. The role of the diet structure: the labour return as a function of the ratio of the cereal food to the domestic animal food in the diet, $\varepsilon_g/\varepsilon_d$, for tillage with hand tools (solid) and with ard (dashed). Dotted line shows the local subsistence carrying capacity $K_s$. The cereal yield and the total contribution of cereals and domestic animal food to the diet are assumed to be constant at $Y = 700$ kg/(ha year) and $\varepsilon_g + \varepsilon_d = 0.8$, respectively.

The variation of the labour return and the local subsistence carrying capacity with the relative contribution of cereals to the diet is shown in Figure 4 assuming a constant cereal yield of $Y = 700$ kg/ha/year and a constant contribution of the wild animal food to the diet, $\varepsilon_w = 0.2$. Solid and dashed lines show the dependencies obtained under the land cultivation with hand tools and with the ard, respectively. A significant constraint that arises if hand tools are used is that a family plot can be tilled in less than 60 person-day/family only for small contributions of cereals to the diet, $\varepsilon_g/\varepsilon_d < 0.4$. An advantage of a diet with a small fraction of cereals, which could be attractive at early stages of the development of farming, is that the labour return is higher when the cereal fraction is lower. For $\varepsilon_g/\varepsilon_d < 0.4$, the labour return exceeds 10 even if hand-tools are used.

With the ard, the tillage takes less than 10 person-day/family for $\varepsilon_g/\varepsilon_d < 4$, and the labour return is exceeds 10 for any reasonable fraction of cereals in the diet.

The size of the exploitation territory remains reasonable across a large part of the range shown in Figure 4, with $0 < D_1 < 0.7$ km, $6.9 > D_2 > 1.5$ km and $7.0 > D_3 > 1.6$ km for $0 < \varepsilon_g/\varepsilon_d < 3$ and a settlement of 2 ha in area with 50 inhabitants. A larger settlement has larger exploitation territory, with $0 < D_1 < 1.5$ km, $3.1 < D_2 < 3.4$ km and $3.2 < D_3 < 3.5$ km for $0 < \varepsilon_g/\varepsilon_d < 3$. The sense of the inequalities for $D_2$ and $D_3$ changes as compared to the smaller settlement because the radii of the grazing and fodder areas are larger while the zone area increases quadratically with its radius.

An increase in the fraction of cereal products (larger $\varepsilon_g/\varepsilon_d$) beyond the equal split, $\varepsilon_g \approx \varepsilon_d$, affects the labour return rather weakly. The labour return varies from 8 to 6 for hand-tillage and from 12 and 10 under ard-tillage as $\varepsilon_g/\varepsilon_d$ increases from 1 to 10. (However, values of $\varepsilon_g/\varepsilon_d$ above about 0.5 do not appear to be practical with hand-tilling because of the time constraints noted above.) Thus, the diet structure, at least after the introduction of the ard, is flexible in this sense as long as the contribution of cereals is large enough, allowing much room for change without any strong effect on the amount of labour required to support it. The change in the labour efficiency is relatively weak mainly because changes in $\varepsilon_g/\varepsilon_d$ lead to a seasonal redistribution of the labour cost between collecting winter fodder and tilling the land and harvesting. Thus, a diet dominated by cereals permits a change of labour resources with little effect on the labour efficiency in case of poor or even failed harvest or any other hazard in food production.

The proportion of cereal food affects noticeably the local subsistence carrying capacity since a larger fraction of domestic animal products makes the economy more land-extensive through the demand for grazing and fodder lands. The magnitude of $K_s$ increases slightly faster than linearly with $\varepsilon_g/\varepsilon_d$ as stronger reliance on cereals means smaller exploitation area.

These calculations confirm that changing the diet can hardly help to remove the labour-cost and time bottlenecks in the soil preparation for sowing: only when the contribution of the cereals is less than half of that from domestic animal products, can two workers till the fields of a family of six in less than 30 days. A diet with similar contributions of cereal and domestic animal products is only possible if the ard replaces hand tools in the land tillage. On the other hand, a diet dominated by cereals (where possible) is rather flexible, and can be adjusted widely without much effect on the labour return. This observation may be relevant to discussions of the risks involved in growing cereals: if the harvest is poor but the (reduced) dominance on cereals can still be maintained (e.g., because of a stored grain), a stronger reliance on animal products does not affect the labour efficiency much, but rather requires a seasonal redistribution of the workload.

To make this point clearer, consider another trajectory in the parameter space that may help to clarify possible risk management strategies associated with arable agriculture. Figure 5 shows the variation of the labour return and the relative fraction of cereals in the diet with the cereal yield, where we assume that the relative contribution of cereals to the diet is proportional to the cereal yield, $\varepsilon_g = 0.4Y/700$ kg/ha/year, keeping the total contribution of domestic products constant, $\varepsilon_g + \varepsilon_d = 0.8$. The crops area is then independent of the cereal yield remaining equal to 0.26 ha/person. This scenario is supposed to model the reaction to a failed harvest or a possible diet evolution as the cereal yield increases systematically with time (e.g., because of the selection of cereal varieties).

With the cereal fraction increasing together with the cereal yield, the labour return is significantly higher, and variation with the yield weaker, than in the case of a fixed diet illustrated in Figure 4. This version of the subsistence strategy is apparently advantageous as it both maximizes the labour return and provides flexibility in terms of the redistribution of resources between growing of crops and animal husbandry. As mentioned above, this strategy may also help to offset the damage of a failed harvest.

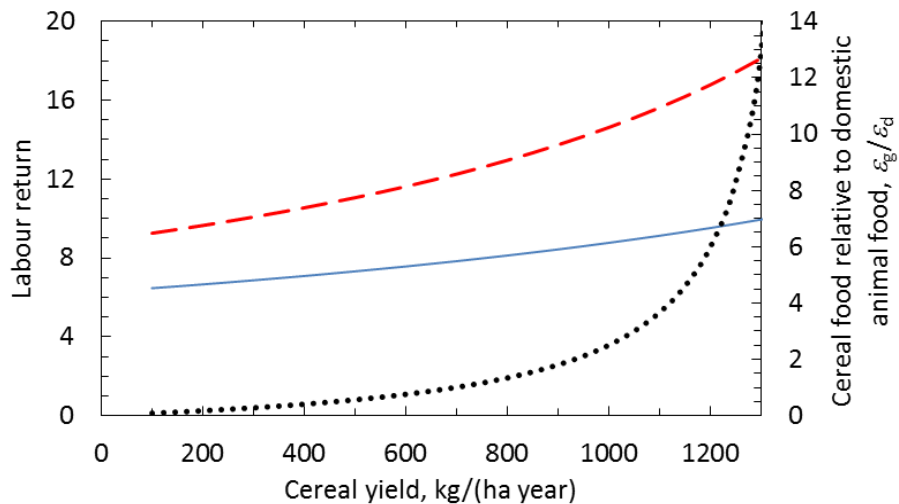

Figure 5. The effect of the cereal yield under a different diet structure on the annual labour return with hand tools (solid) and ard tillage (dashed), and the ratio of the contributions of cereals and domestic animal products to the diet, $\varepsilon_g/\varepsilon_d$ (dotted). The contribution of cereals is assumed to be proportional to the cereal yield, $\varepsilon_g = 0.4Y/700$ kg/(ha year), and the total contribution of domestic foods to the total diet is kept constant, $\varepsilon_g + \varepsilon_d = 0.8$.

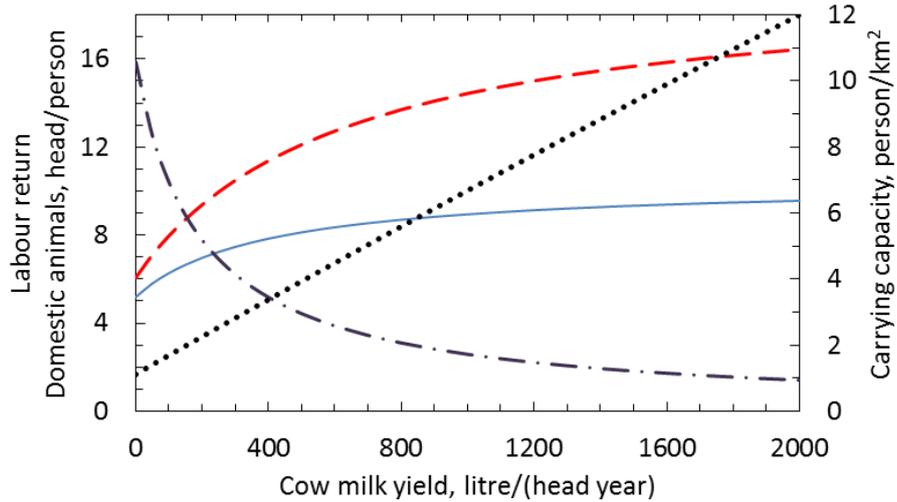

Figure 6. The effect of the milk yield on the labour return (solid: hand-tool tillage; dashed: ard tillage), the local subsistence carrying capacity (dotted) and the per capita number of domestic animals required to satisfy dietary requirements (dash-dotted).

### 8.3.3 The role of dairy products

Dairy products appear to have been a part of the European human diet since the Early Neolithic. However, previous palaeoeconomy analyses rarely, if ever, included dairy products. The general attitude felt in the literature is that they are an attractive but optional addition rather than an essential component of the diet. Based on our calculations, we argue that milk and dairy products could be an essential component of the diet, providing an opportunity to reduce labour costs.

Figure 6 illustrates the role of the dairy products. The results shown are obtained by increasing the cow and caprine milk yields together, $y_s = 50\,(y_c/400)^{1/3}$, where both $y_s$ and $y_c$ are measured in litres/head/year. This dependence is chosen exclusively for illustrative purposes to ensure that the range of variation of the caprine milk yield is reasonable as the cow milk yield varies. In particular, the nominal figures $y_s = 50$ litre/head/year for $y_c = 400$ litre/head/year are reproduced, and, at the top end of the range, $y_s = 146$ litre/head/year for $y_c = 10{,}000$ litres/head/year are similar to the modern livestock figures.

Unsurprisingly, increasing milk yield boosts the labour return. What is surprising is that the effect is so significant. As the milk yield increases from zero to 2000 litre/head/year, the efficiency of the hand-tool agriculture grows from 5 to 10, and labour assisted by the ard has the return boosted from 6 to 16. For larger milk yields, the size of the herd required to satisfy the dietary requirements reduces, and hence the grazing and fodder zones become smaller. As a result, the local carrying capacity increases with the milk yield linearly from 1 to 12 persons/km$^2$ as $y_c$ increases from 0 to 2000 litres/head/year and $y_s$ increases simultaneously from 0 to about 90 litres/head/year. We neglect the labour costs of milking, tending the animals, collecting leaf fodder, etc., and this, of course, contributes to the increase in the labour return. Again, these activities can be assigned to the weaker family members: the labour returns quoted here refer to the physically most demanding activities performed by a few physically stronger people.

The effect of the milk yield on the carrying capacity is so strong because the number of domestic animals that need to be kept reduces significantly if their milk is used for food. The dash-dotted curve in Figure 6 shows how rapidly the per capita number of the livestock decreases as the milk yield increases. For $y_c = 0$, an implausibly large herd of 16 heads is required to satisfy the dietary requirements of a single person for the diet structure assumed ($\varepsilon_g/\varepsilon_d/\varepsilon_w = 0.4/0.4/0.2$). We discussed above how such an implausible situation could be avoided, but this stresses once more the importance of dairy products. For $y_c = 400$ litres/head/year, the herd size decreases to about 5.2 heads per capita (1.8 heads of cattle, 1.2 caprines, 1.7 pigs and 0.4 horses per person), still a rather large herd to keep. The rapid decrease continues to 1.4 head/person, comprising 0.5 cattle, 0.3 sheep or goats, 0.5 pigs and 0.1 horses for $y_c = 2000$ litres/head/year. It is clear that even for this productivity of the milk herd, still low by modern standards, there are many opportunities to produce surplus product beyond the subsistence requirements.

To provide another illustration of the importance of dairy products, we note that, if no milk is used at all and the fraction of domestic animal products (then, meat alone) remains equal to $\varepsilon_d = 0.4$, the size of per capita herd increases to $n_a = 16$ head/person for the nominal parameters values, clearly an untenable number. Moreover, the daily consumption of meat from domestic animals alone becomes as high as 500 g/person/day (as compared to 160 g/person/day if milk is used). To put this figure into a simple but relevant context, we note that a fillet steak served in a typical British restaurant weighs 230 grams. It is thus clear that palaeodiet reconstructions with a significant fraction of animal products are inconsistent with the constraints of human physiology and nutrition unless a significant part of the animal food are dairy products.

### 8.3.4 The exploitation territory

The above discussion contains references to the size of the exploitation territory of a settlement in connection with the expectation that the distance to the arable fields should not exceed 5 km and preferably be within 1–2 km of a settlement, whereas the distance to the pasture areas should be within 5–10 km. In this section, we summarize this aspect of our results.

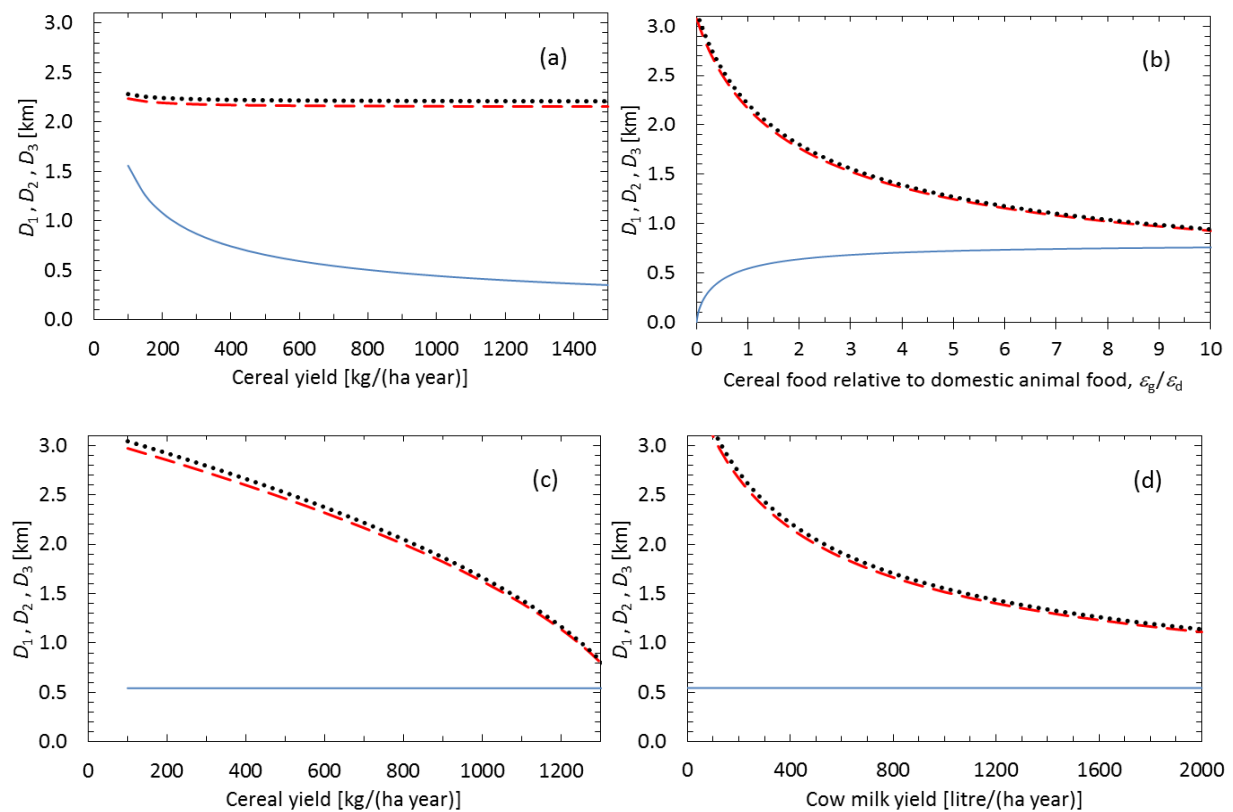

Figure 7. The maximum distances from the settlement border ($R_0 = 80$ m) to the three exploitation zones around a settlement of 2 ha in area with about 50 inhabitants: $D_1$, the field zone (solid); $D_2$, the grazing zone (dashed); and $D_3$, the fodder zone (dotted). The model illustrated in each panel is one of those discussed and illustrated above: (a) Figure 3, (b) Figure 4, (c) Figure 5, and (d) Figure 6.

Figure 7 shows the maximum distances from a settlement border to the field zone ($D_1$), grazing zone ($D_2$) and the fodder zone ($D_3$). For this illustration, we have chosen a typical settlement size of the Early Trypillia, of an area $A_0 = 2$ ( ha with 50 people (Table 2). ). Each panel of Figure 7 corresponds to one of the models discussed above and illustrated in Figures Figure 3–Figure 6. The maximum distance to the fields $D_1$ is close to 0.5 km in all cases except for extremely low cereal yields (Panel a) or extremely high fraction of cereals in the diet (Panel b), but even then it does not exceed 1.5–0.8 km. The distance to the grazing area, $D_2$, never exceeds 3 km and is smaller than 2 km under rather realistic choices of parameters. The distance to the fodder zone, $D_3$, differs from $D_2$ insignificantly because the radius of this zone is large, and hence even a narrow annulus can have a substantial area.

The situation is not that simple for larger settlements. Assuming, for the sake of argument, that the population density is independent of the population size (375 m² of the total settlement area per person), a settlement of an area 10 ha has about 270 people. With the nominal values of parameters of Table 8, the outer radii of the three zones, $D_1 = 1.2$ km, $D_2 = 4.8$ km and $D_3 = 5.0$ km, are approaching the maximum acceptable values. A 40-ha settlement (1100 people) is only marginally sustainable with $D_1 = 2.4$ km, $D_2 = 9.7$ km and $D_3 = 9.9$ km. Of course, optimisation of the subsistence strategy by changing the diet (perhaps only slightly) or a higher cereal or meal yield, to mention just a few options, can make a 40-ha village a viable option. We note that the amount of fallow land adopted (twice the area of the fields under direct cultivation, $\delta_f = 2$) might be unrealistically small as it implies a triennial fallow. Early agricultural systems could use longer fallow intervals; ethnographic data suggest that fallow length of 8–15 years is not unusual (Styger and Fernandes 2006). Longer fallow would obviously result in larger exploited land area. Notably, the median size of the Trypillia settlements given in Table 2 does not exceed 8.4 ha. It is clear, however, that significantly larger settlements would need a fundamental change in the organization of their food supplies, and the division of labour and occupation, with ensuing increased social complexity is an obvious option.

Figure 8 shows the variation of the maximum distance to the field zone from a settlement boundary with the size of the fallow area relative to the cropped area for several typical settlement sizes. It is noteworthy that the distances to the grazing and fodder zones do not change as $\delta_f$ varies since the larger fallow land is used for pasture, so that the size of the grazing area reduces as the fallow area increases.

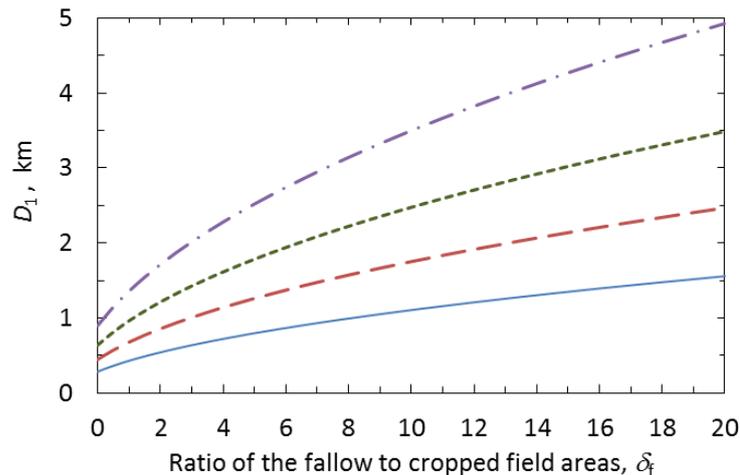

Figure 8. The dependence of $D_1$, the maximum the distance from a settlement boundary to the field area (see Figure 2) for settlements of various areas and populations: 2 ha, 50 people (solid), 5 ha, 130 people (long-dashed), 10 ha, 270 people (short-dashed) and 15 ha, 400 people (dash-dotted).

The distance to the field zone increases with the fallow ratio at a modest rate (roughly, as the square root of $\delta_f$). The field zone is within 1–2 km of a village only if the length of the fallow is not too large: $D_1 < 1.5$ km for $\delta_f < 20$ for a settlement of 2 ha in area but only for $\delta_f < 7.5$ around a 5 ha settlement. The fields of a bigger settlement of 10 ha are within this distance only for $\delta_f < 3$; for a still larger settlement of 20 ha, the maximum acceptable value of $\delta_f$ is the marginal 1.5. Of course, these figures would be smaller for a higher cereal yield, with $D_1$ decreasing roughly in inverse proportion to the square root of the yield. However, this illustrates once more that settlements of more than a few tens of hectares, with more than a few hundred people, are likely to function differently from smaller villages as the need to import food from satellite farming villages rapidly increases with the size of the settlement.

## 8.4 Surplus food production

The above estimates present an overall economic picture of farming based on the immediate dietary requirements of the population. There is another aspect of this picture that we have touched upon only in passing: the risks of agricultural production mainly associated with failed crops (Halstead 2004). The diversification of the domesticated plants and livestock, storage of emergency reserves and wider use of wild resources are among the strategies used to mitigate this risk. However, the storage for emergencies obviously requires some surplus of food to be produced implying higher labour costs. The opportunity to produce a surplus product can also profoundly affect the economic behaviour of the farmer. If a surplus product can be, and indeed is, produced beyond the needs of the farmers and their families, the importance of transportation and communication greatly increases, as the surplus produce needs to be transported to the consumer on a regular basis. This makes it more important for the farm to be located conveniently with respect to (most often, close to) transportation routes, of which waterways are most obvious. In turn, this makes isolated hamlets a less attractive option for a farmer to occupy, thereby facilitating the agglomeration and clustering of the population.

In the discussion above, we have identified a direct route to a surplus food production via the use of dairy products: by providing a significant food resource that requires relatively little labour investment from the physically fit family members, it provides an opportunity to redirect the resources to producing surplus product in any branch of agriculture. We shall explore these opportunities in another publication.

## 9. The lifetime of a farming settlement

### 9.1 Archaeological evidence

There are many indications that most Trypillia settlements had a relatively short lifetime of less than 100 years. Most of the settlements have a single-layer stratigraphy. Tells are found only in the Carpathian piedmont areas, and even there only isolated phases and stages are represented in the excavation finds, often separated by significant gaps. This is also true of the multi-layered sites discovered in the eastern part of the CTU area, where material finds are restricted to 2–3 phases. For example, the largest settlements, such as Talianky and Maydanetske, belong to a limited part of the same stage, CI (Smaglii and Videiko 1990; Ryzhov 1990).

There were several attempts to estimate the Trypillia settlement lifetime, converging to 50–100 years (e.g., Krutz 1989; Markevich 1981). These estimates were based on archaeological dating and pottery typology, together with $^{14}$C and archaeomagnetic dating. For example, Ryzhov (1990) identified distinct phases in the development of the Trypillia sites in the Dnieper–Southern Bug interfluves in the fourth millennium BC. The types of painted pottery found there suggest up to five development phases belonging to Stage BII and four, to CI, nine phases altogether. According to archaeomagnetic dating, the overall duration of these phases is 500–600 years (Telegin 1985, pp. 11–17). The author of these archaeomagnetic measurements, G. F. Zagnii (private communication) suggests that their accuracy is 25–50 years, sufficiently high for our purposes; for comparison, recent archaeomagnetic studies of Neolithic sites in Bulgaria (Jordonova et al., 2004) and Greece (Aidona and Kondopoulou, 2012) report accuracies of up to ±70 and ±85 years, respectively (from the 95% range of dates). (The accuracy of the archaeological dates obtained from $^{14}$C measurements is as yet insufficient to make them useful in this discussion.) The average duration of a single phase, which can be identified with the settlement lifetime, follows as 50–70 years. However, the stratigraphic structure within a single phase (e.g., Maydanetske – Shmaglii and Videiko 2001–2002) suggests that at some sites the lifetime could be somewhat longer but never exceeding 80–150 years. In the vast majority of cases, a repeated occupation of a given site, if it happened, occurred with prolonged periods of abandonment, often of 200–500 years long.

## 9.2 A depleted resources model

From the available archaeological and agricultural evidence, it is possible to estimate the maximum lifetime of a farming settlement if it is limited by the decreasing soil fertility alone. We assume a settlement has a fraction $f$ of fallow land. Thus, at any time, a plot is either being farmed, and so has decreasing fertility, or it is fallow and then its fertility is recovering. Denote $\delta_F$ the ratio of the fallow to cropped areas. Let $T_R$ be the recovery time scale of the soil fertility and $T_D$ the fertility depletion time. In a depletion phase (i.e., when a field is being farmed) we have a decreasing content of soil nutrients, which can be described as an instantaneous reduction in the potential yield,

$$Y(t) = Y_0 \exp(-t/T_D),$$

where $Y_D$ denotes the crop yield at a time $t$ in the cultivation phase that starts at $t = 0$, and $Y_0$ is the starting yield (e.g., that of the of virgin land). When discussing the Sanborn data above, we used linear fits to the yield variation with the time span after the stat of the cultivation, which proves to be sufficient over relatively short periods of order 30 years. On longer timescales, the yield is likely to decrease exponentially with time as adopted here, assuming that a constant *fraction* (rather than the *amount*) of nutrients is extracted annually from the soil by the crop plants.

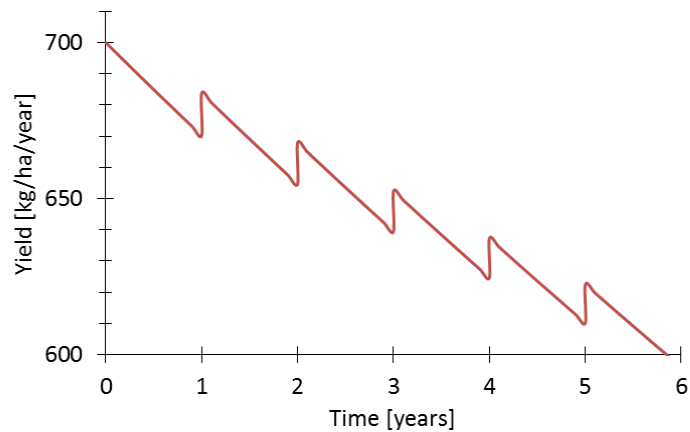

Figure 9. An illustration of the cereal yield changes in a fallow system with the initial yield $Y_0 = 700$ kg/ha/year, the ratio of fallow to cultivated field areas $\delta_F = 2$ (so that any given plot is used for the crops for one year and then stays fallow for two years), the fertility depletion time scale $T_D = 23$ years and the fertility recuperation time scale $T_R = 100$ years. Note that the time plotted includes only the periods of cultivation. (The fallow periods correspond to the step-increases of yield after each year of cultivation.) For total elapsed time, these numbers should therefore be multiplied by $(1 + \delta_F) = 3$.

Suppose that the plot is farmed for a period $t_1$ and then left fallow for a period $t_2 = \delta_f t_1$ while one of the other $\delta_f$ plots is cultivated. The recovery of the potential yield of the fallow field is then described by

$$Y(t) = Y(t_1) \exp \frac{t-t_1}{T_R}, \qquad t_1 < t \leq t_1(1+\delta_f).$$

A complete recovery of a fallow field is not waited for, simply as it takes too long. Rather, a plot is cropped again after the full rotation, at $t = (1 + \delta_f)t_1$, and the cycle repeats again and again. The resulting variation of the yield from the whole land area containing all the plots involved is shown in Figure 9. The cycle is repeated until the yield reduces to a level $Y_m$ too low to be useful, and then the whole site is abandoned and a new settlement location is sought. The average yield (with the saw-tooth changes smoothed-out) is then given by

$$Y(t) = Y_0 \exp\left[-\frac{t}{t_1(1+\delta_f)}\left(\frac{1}{T_D} - \frac{\delta_f}{T_R}\right)\right].$$

The land will be abandoned when abandoned at the time $T$ such that $Y(T) = Y_m$, and the settlement lifetime $T$ then follows as

$$T = t_1(1+\delta_f)\frac{T_D T_R}{T_R - \delta_f T_D} \ln\frac{Y_0}{Y_m}.$$

From the fits to the Sanborn data discussed above, the average half-life of a plot of land (i.e., the average time for the yield to halve) is approximately $\tau_u = 17$ years for unmanured plots and $\tau_m = 28$ years where manure fertilizer was applied. This gives the combined half-life of $T_D = [(1-f_m)/\tau_u + f_m/\tau_m]^{-1} \approx 20$ years, for the fraction of manured fields $f_m = 0.4$. We assume a recovery time of $T_R = 100$ years (e.g., Boserup 1965) and adopt $\delta_f = 2$ and $Y_m = 250$ kg/ha/year, which corresponds to the minimum acceptable labour return with hand tillage of three in Figure 3. For $Y_0 = 700$ kg/ha/year and $t_1 = 1$ year, we obtain $T \approx 130$ years as the settlement lifetime in this model. This estimate is rather sensitive to the amount of land kept fallow but only weakly to the minimum yield leading to the settlement to be abandoned. For example, for $\delta_f = 3$ and other parameters unchanged, we obtain $T \approx 300$ years.

## 10. Conclusions and discussion

From the very beginning of its evolution, the CTU possessed a developed agricultural technology with a wide spectrum of domesticated plants and animals. We present palaeoeconomy reconstructions of pre-modern agriculture selecting, wherever required, features specific for the CTU, and paying special attention to the self-consistency of all the elements of the model within the constraints provided by the archaeological, environmental and technological evidence available. With full appreciation of the tentative and approximate nature of any estimates of this kind, our calculations firmly demonstrate the sustainability of the CTU agriculture. Our models include several equally important elements. We start with the calorific content of the palaeodiet suggested by archaeological data, stable isotope analyses of human remains, and palynology studies in the area. We allow for all known domestic and wildlife elements of the diet and provide plausible estimates of the pre-modern yield of ancient cereal varieties and its dependence on the rainfall and duration of continuous land cultivation. Importantly, we pay proper attention to the labour costs of various seasonal parts of the agricultural cycle, not only for an individual but also for the farmer's family (with its majority of weak and young members not capable of hard physical labour); this was rarely, if ever, done systematically in the earlier studies of pre-modern agri-culture. Finally, we put our results into the context of the exploitation territory and catchment analysis to translate the subsistence needs and strategy of an individual to those of settlements of various sizes. Many (but not all) aspects of the economy are conveniently summarised in terms of the labour return, the ratio of the amount of food energy produced to the energy spent or, equivalently, the total amount of labourer-time available to the working time. Another important aspect of the agricultural activities is the relation of the labour productivity to the time available to seasonal agricultural activities. Of those, the land preparation for sowing causes the strongest time stress. We address this aspect of the problem using the published results of experiments on tillage, reaping, threshing and winnowing using primitive tools and/or traditional techniques.

The simplest subsistence strategy, based on a complex of cereals, domestic and wild animal products, with fallow cropping, appears to be capable of supporting an isolated, relatively small farming community of 100–300 people even without recourse to technological improvements such as the use of manure fertiliser. The most important factor limiting the size of such a community is the labour productivity and the labour cost of land cultivation with hand tools. The time stress at the crop sowing time can be relieved by reducing the fraction of cereals in the diet to about 25% in terms of calorific content. Reduction in the soil fertility with time, estimated here from the continuous agricultural experiment on virgin land at Sanborn (Missouri, USA), suggests that soil fertility around such a settlement would be depleted within 60–100 years even with a fallow system. This factor can determine the lifetime of a farming village. Such settlements are typical of earliest Trypillia Stage A.

A larger settlement of several hundred people could function in isolation, and with a larger fraction of cereals in the diet, only with technological innovations; for example, the use of manure fertiliser and, most importantly, the use of the ard for land tilling. The ard relieves radically the extreme time pressure at the time of soil preparation for sowing. There is archaeological evidence for the use of ard from the Trypillia Stage BI. Another constraint on the settlement size arises from the fact that animal husbandry is land-extensive, and the distance to the grazing area increases very rapidly with the settlement size. It appears that very large settlements of a few hundred hectares in area could function only if supported by satellite farming villages. In turn, this implies division of labour, sufficiently complex social relations, stable exchange channels, etc.: altogether, a proto-urban character of such settlements.

Arable agriculture is more labour expensive and involves stronger seasonal time stress than animal husbandry. However, variations in the labour return with the fraction of cereals in the diet indicate that a diet dominated by cereals is more flexible in the sense that labour redistribution between obtaining food from cereals and domestic animals does not affect the labour return significantly but leads to a seasonal redistribution of the labour costs. This feature can be relevant to the mitigation of the risk of failed crops: when cereals dominate in the diet, applying more effort to the livestock is easy in this respect. Another ways to counter the risk is the use of the manure fertiliser as it significantly reduces the yield variability. We quantify this using the Sanborn experimental data.

Yet another strategy to handle the agricultural risks is the storage of an annual supply of grain to be used when the harvest is low. Typical labour returns of order $\eta = 6$–$8$ if using hand tools for the tillage and $\eta = 10$ for the ard tillage imply that keeping such a storage is indeed possible. In a family of six with two members fit for hard agricultural labour (so that each of the workers feeds three people), the minimum labour return required for immediate subsistence is $\eta = 3$. Any effort beyond this figure can be used to produce a surplus, part of which can be stored as insurance.

Even when the insurance grain storage has been laid out, there is sufficient reserve in the labour return to produce surplus food that can be exchanged or traded externally. However, the tillage bottleneck prevents significant surplus grain being produced unless the ard is used to till the land. Thus, exchange networks, labour division, etc., can indeed be expected to develop starting from the middle CTU stages.

The significant fraction of cattle and horses in the CTU faunal assemblages and osteometric evidence of their use for traction suggest that agricultural activities involved more than one (extended) family to justify the costs of maintaining animals for anything other than food (Halstead 1996). Reducing the fraction of cattle in the herd from the nominal figure 0.35 to 0.25 (say, at the expense of the caprines), and simultaneously increasing the fraction of milked animals from 0.5 to 0.7 to keep the number of milking cows the same, reduces the annual subsistence labour cost from 47 to 22 person-day/person/year and the labour return increases from 7.8 to 8.3. This example (illustrative and not necessarily realistic) clearly demonstrates the costs of keeping traction animals and stresses advantages of cooperation between farmers who need animal traction only for limited periods in the seasonal agricultural cycle. A heard of thirty cattle may be a minimum herd size for reproductive maintenance of a herd (Bogucki 1982, p. 109; Glass 1991, p. 28). The need to combine the resources of several farming households may be another factor that determines the minimum size of an isolated farming village.

It is tempting to apply the palaeoeconomy model to later stages of the CTU development and to larger settlements. However, larger settlements are rare and, hence each of them is special. Therefore, such an application should be based on careful analysis of the landscape and environment at the giant CTU settlements. Gaydarska (2003) has started such work for Maydanetske. In addition, quantitative analysis of connections between CTU sites, e.g., suggested by pottery typology, is required to assess the intensity of exchange networks.

## Acknowledgements

This work was supported by the EC FP6 project FEPRE and the Leverhulme Trust Research Grant F/00 125/AD.